\documentclass[conference,compsoc]{IEEEtran}

\ifCLASSOPTIONcompsoc
  \usepackage[nocompress]{cite}
\else
  \usepackage{cite}
\fi

\usepackage{nicefrac}
\usepackage{siunitx}
\usepackage{array,framed}
\usepackage{booktabs}
\usepackage{
  color,
  float,
  epsfig,
  wrapfig,
  graphics,
  graphicx,
  subcaption
}

\usepackage{textcomp,amssymb}
\usepackage{setspace}
\usepackage{latexsym,fancyhdr,url}
\usepackage{enumerate}
\usepackage{enumitem}
\usepackage[most]{tcolorbox}

\usepackage{algorithm}
\usepackage[noend]{algpseudocode}
\usepackage[symbol]{footmisc}
\renewcommand{\thefootnote}{\arabic{footnote}}
\usepackage[numbers,sort]{natbib}

\usepackage{graphics}
\usepackage{xparse} 
\usepackage{xspace}
\usepackage{multirow}
\usepackage{csvsimple}
\usepackage{balance}
\usepackage{pifont}
\usepackage{svg}

\usepackage{
  tikz,
  pgfplots,
  pgfplotstable
}
\usepackage{xcolor}
\definecolor{darkgreen}{rgb}{0.0, 0.5, 0.0} 
\usepackage[hidelinks]{hyperref}
 \hypersetup{
     colorlinks=true,
     citecolor=darkgreen,     
     linkcolor=red!70!black,       
     urlcolor=blue!70!black        
 }
\usepackage[capitalise,nameinlink, noabbrev]{cleveref}

\newcommand{\email}[1]{{\rm\textsf{\href{mailto:#1}{#1}}}}

\definecolor{darkviolet}{HTML}{9400D3}

\usetikzlibrary{
  shapes.geometric,
  arrows,
  external,
  pgfplots.groupplots,
  matrix
}

\pgfplotsset{compat=1.9}

\usepackage{mathtools}

\DeclareMathAlphabet{\mathcal}{OMS}{cmsy}{m}{n}

\usepackage{svg}

\DeclareGraphicsExtensions{%
    .png,.PNG,%
    .pdf,.PDF,%
    .jpg,.mps,.jpeg,.jbig2,.jb2,.JPG,.JPEG,.JBIG2,.JB2}

\usepackage{xparse}
\newcommand{\bnm}{\begin{newmath}}
\newcommand{\enm}{\end{newmath}}

\newcommand{\bea}{\begin{eqnarray*}}%
\newcommand{\eea}{\end{eqnarray*}}%

\newcommand{\bne}{\begin{newequation}}
\newcommand{\ene}{\end{newequation}}

\newcommand{\bal}{\begin{newalign}}
\newcommand{\eal}{\end{newalign}}

\newenvironment{newalign}{\begin{align}%
\setlength{\abovedisplayskip}{4pt}%
\setlength{\belowdisplayskip}{4pt}%
\setlength{\abovedisplayshortskip}{6pt}%
\setlength{\belowdisplayshortskip}{6pt} }{\end{align}}

\newenvironment{newmath}{\begin{displaymath}%
\setlength{\abovedisplayskip}{4pt}%
\setlength{\belowdisplayskip}{4pt}%
\setlength{\abovedisplayshortskip}{6pt}%
\setlength{\belowdisplayshortskip}{6pt} }{\end{displaymath}}

\newenvironment{newequation}{\begin{equation}%
\setlength{\abovedisplayskip}{4pt}%
\setlength{\belowdisplayskip}{4pt}%
\setlength{\abovedisplayshortskip}{6pt}%
\setlength{\belowdisplayshortskip}{6pt} }{\end{equation}}

\newcounter{ctr}

\newcounter{mytable}
\def\mytable{\begin{centering}\refstepcounter{mytable}}
\def\endmytable{\end{centering}}

\newcounter{myfig}
\def\myfig{\begin{centering}\refstepcounter{myfig}}
\def\endmyfig{\end{centering}}

\newlength{\saveparindent}
\newlength{\saveparskip}
\newcommand{\E}{{\rm I\kern-.3em E}}

\renewcommand{\eqref}[1]{\mbox{Equation~(\ref{#1})}}

\newcommand{\mV}{m\kern-1pt V\xspace}
\newcommand{\Vpp}{V\kern-1pt \textsubscript{pp}\xspace}

\def \part {part}

\renewcommand{\paragraph}[1]{\vspace*{6pt}\noindent\textbf{#1}\;}

\def \blackslug{\hbox{\hskip 1pt \vrule width 4pt height 8pt
    depth 1.5pt \hskip 1pt}}
\def \qed{\quad\blackslug\lower 8.5pt\null\par}

\newcounter{mynote}[section]

\newcommand\ignore[1]{}

\newcounter{rcnote}[section]

\newcounter{mrnote}[section]

\newcounter{fknote}[section]

\newcounter{anote}[section]

\DeclareMathSymbol{\mlq}{\mathord}{operators}{``}
\DeclareMathSymbol{\mrq}{\mathord}{operators}{`'}

\newcommand{\rhf}[2]{R_{f, \gamma}}

\DeclareDocumentCommand{\edist}{o o}{
  \ensuremath{
    \IfNoValueTF{#1}{{d}}{{\sf d}(#1,#2)}
  }
}

\newcommand{\olrk}[1]{\ifx\nursymbol#1\else\!\!\mskip4.5mu plus 0.5mu\left(\mskip0.5mu plus0.5mu #1\mskip1.5mu plus0.5mu \right)\fi}

\NewDocumentCommand{\indseq}{ O{1} O{r} }{{#1}\ldots {#2}}

\makeatletter
\newcommand{\linebreakand}{%
  \end{@IEEEauthorhalign}
  \hfill\mbox{}\par
  \mbox{}\hfill\begin{@IEEEauthorhalign}
}
\makeatother

\newcommand{\parhead}[1]{\vspace{3pt plus 1pt minus 3pt}\par\noindent\textbf{#1}\hspace{.4em plus .2em minus .2em}}

\definecolor{byzantium}{rgb}{0.44, 0.16, 0.39}

\newif\ifshowcomments
\showcommentstrue 

\begin{document}
\fancyhead{}
\def\thetitle{Chypnosis: Undervolting-based Static Side-channel Attacks} 

\title{\thetitle}


\author{\IEEEauthorblockN{Kyle Mitard}
\IEEEauthorblockA{Worcester Polytechnic Institute\\
\small\email{krmitard@wpi.edu}}
\and
\IEEEauthorblockN{Saleh Khalaj Monfared}
\IEEEauthorblockA{Worcester Polytechnic Institute\\
\small\email{skmonfared@wpi.edu}}
\and
\IEEEauthorblockN{Fatemeh Khojasteh Dana}
\IEEEauthorblockA{Worcester Polytechnic Institute\\
\small\email{fdana@wpi.edu}}
\linebreakand
\IEEEauthorblockN{Robert Dumitru}
\IEEEauthorblockA{Ruhr University Bochum \&\\ The University of Adelaide\\
\small\email{robert.dumitru@adelaide.edu.au}}
\and
\IEEEauthorblockN{Yuval Yarom}
\IEEEauthorblockA{Ruhr University Bochum\\
\small\email{yuval.yarom@rub.de}}
\and
\IEEEauthorblockN{Shahin Tajik}
\IEEEauthorblockA{Worcester Polytechnic Institute\\
\small\email{stajik@wpi.edu}}}







\date{}

\maketitle
\renewcommand{\thefootnote}{}
\footnotetext{\textcolor{blue}{This article has been accepted for publication in \textit{IEEE Symposium on Security and Privacy 2026}.}}
\renewcommand{\thefootnote}{\arabic{footnote}}

\thispagestyle{plain} 
\pagestyle{plain}

\begin{abstract}

Static side-channel analysis attacks, which rely on a stopped clock to extract sensitive information, pose a growing threat to embedded systems' security. 
To protect against such attacks, several proposed defenses aim to detect unexpected variations in the clock signal and clear sensitive states.
In this work, we present \emph{Chypnosis}, an undervolting attack technique that indirectly stops the target circuit clock, while retaining stored data.
Crucially, Chypnosis also blocks the state clearing stage of prior defenses, allowing recovery of secret information even in their presence.
However, basic undervolting is not sufficient in the presence of voltage sensors designed to handle fault injection via voltage tampering.
To overcome such defenses, we observe that rapidly dropping the supply voltage can disable the response mechanism of voltage sensor systems.
We implement Chypnosis on various FPGAs, demonstrating the successful bypass of their sensors, both in the form of soft and hard intellectual property (IP) cores.
To highlight the real-world applicability of Chypnosis, we show that the alert handler of the OpenTitan root-of-trust, responsible for providing hardware responses to threats, can be bypassed. 
Furthermore, we demonstrate that by combining Chypnosis with static side-channel analysis techniques, namely laser logic state imaging (LLSI) and impedance analysis (IA), we can extract sensitive information from a side-channel protected cryptographic module used in OpenTitan, even in the presence of established clock and voltage sensors.
Finally, we propose and implement an improvement to an established FPGA-compatible clock detection countermeasure, and we validate its resilience against Chypnosis.

\end{abstract}

\section{Introduction}\label{sec:intro}
Physical side-channel attacks (SCA) can undermine the security of cryptographic implementations on integrated circuits (ICs).
These attacks typically exploit the inevitable influence of data transitions during computation on current consumption or voltage drop on a chip. 
Dynamic side-channel attacks, such as power~\cite{kocher1999differential} and electromagnetic analysis~\cite{agrawal2002side}, can exploit such data transitions and recover the secret from the chip.
Recently, however, static physical SCA attacks have been gaining attention, in which adversaries can extract static data stored in memories, such as Flip-Flops (FFs). 
Examples of such static attacks include static power analysis~\cite{moradi2014side}, Laser Logic State Imaging (LLSI)~\cite{krachenfels2021real}, Impedance Analysis (IA)~\cite{monfared2023leakyohm}, and Thermal Laser Stimulation (TLS)~\cite{krachenfels2021automatic}. 

Such static attacks require some level of tampering with the clock and voltage of the target chip.
First, the attacker must freeze the circuit's state by halting its clock, because the time required for recovering the static data stored in registers is significantly longer than the clock period~\cite{krachenfels2021real,monfared2023leakyohm}.
Second, in a subclass of these static attacks known as backscatter attacks (e.g., LLSI and IA), the adversary must modulate the voltage supplying the chip to produce a detectable modulated reflection during laser or microwave stimulation.
In these attacks, the adversary stimulates the chip using external signals (e.g., near-infrared laser beams for LLSI or microwave radiation for IA) and measures the modulated reflections to infer the internal circuit state or memory contents. Due to their active nature, static backscatter attacks often achieve a higher Signal-to-Noise Ratio (SNR) and, in some cases, can extract secrets with a single trace, rendering data randomization techniques such as masking ineffective~\cite{krachenfels2021real,monfared2023leakyohm}.

\begin{figure}[t]
    \centering
    \includegraphics[width=0.45\textwidth]{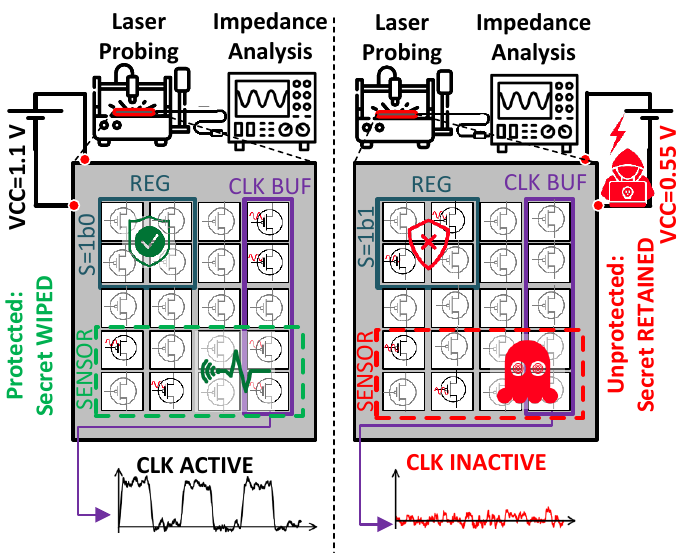}
    \caption{High-level overview of Chypnosis Attacks}
    \label{fig:hypno_hl}
\end{figure}

Consequently, detecting clock or voltage tampering and then responding by wiping the sensitive data is a promising countermeasure against these static attacks. 
Various clock and voltage sensors, both as soft intellectual property (IP) cores~\cite{dumitru2025borrowed,farheen2022twofold} and hard IP cores~\cite{peterson2017developing,microchip_security}, have been developed for this purpose.
Moreover, specialized sensors targeting specific threat vectors, such as voltage glitching~\cite{zick2013sensing,provelengios2021mitigating} and laser probing~\cite{monfared2024laserescape,zhang2023laser,tajik2017pufmon}, can also serve as effective countermeasures against backscatter attacks.
Additionally, the assumption that an attacker can access and manipulate the system clock is often unrealistic in real-world scenarios. Many secure ICs rely on internal clock sources~\cite{microchip_security} for cryptographic operations, making external clock control extremely difficult for an adversary.

Driven by these observations, in this paper, we ask the following  questions:
\emph{(1) Is it possible to halt the system’s clock without tampering with its source? (2) Can we perform static side-channel attacks without triggering clock and voltage sensors?}

\subsection{Our Contribution}
In this work, we positively answer both questions. 
We introduce \textit{Chypnosis} attacks, in which an adversary bypasses clock and voltage sensors, places a chip in hibernation while retaining its data. 
In this condition, the adversary can perform a static side-channel attack to recover the retained data. 
Our attack exploits the observation that rapidly lowering the supply voltage below nominal thresholds induces a brownout condition, where logic components (e.g., state machines and clock buffers) cease switching, but volatile memory elements, such as SRAM and flip-flops, continue to retain data. \cref{fig:hypno_hl} presents an abstract overview of our attack. 

We perform Chypnosis on both SRAM-based and Flash-based FPGAs fabricated using 28\,nm processes. 
First, we conduct extensive experiments to determine the voltage thresholds required to induce hibernation at various clock frequencies. 
Next, we demonstrate that entering this brownout condition effectively halts the clock and suppresses switching activity without requiring direct control over the clock source. 
We confirm this behavior through electrical and optical measurements.
We further show that a switching freeze caused by a rapid voltage drop disables the response circuitry of clock, voltage, and brownout detection (BOD) sensors on FPGAs, irrespective of being implemented as soft IPs~\cite{dumitru2025borrowed,farheen2022twofold} or hard IPs~\cite{microchip_security}.
Hence, the sensitive data is not wiped and remains on the chip.

To demonstrate Chypnosis's effectiveness in practical scenarios, we successfully attack the FPGA implementation of OpenTitan Earl Grey and bypass its \textit{alert handler}~\cite{opentitan}, which is responsible for providing hardware responses to threats.
Finally, we demonstrate that even in the brownout state, it is possible to delicately modulate the supply voltage without crashing or waking the system, thereby successfully performing LLSI and IA attacks.
We demonstrate the extraction of secret data from a side-channel-protected symmetric cryptographic module, used in OpenTitan, in a single trace.
We conclude by proposing a circuit-based FPGA-compatible countermeasure, which successfully mitigates Chypnosis.
\\

\section{Technical Background}\label{sec:backgraound}

\subsection{Conventional Static SCA Countermeasures}\label{sec:conv_counter}

\subsubsection{Detection-based Countermeasures}

\parhead{\\Clock Sensors.}
The primary requirement for launching a static SCA attack is to stop the system clock.
Hence, if we have a sensor that detects that the clock has been halted for a while, it can trigger a response, such as wiping sensitive data, before an adversary can recover it.
Some secure IC families are equipped with internal clock sensors capable of detecting anomalies in clock behavior~\cite{microchip_security}. However, the specific design details of these sensors are proprietary.
FPGAs also contain internal circuitry, such as phase-locked loops (PLLs), that can detect irregularities in the clock waveform; however, they must be explicitly configured by the user to function as a security sensor. 
Commercial solutions, such as the AMD/Xilinx Security Monitor IP core~\cite{SECMON}, offer built-in clock monitoring capabilities; however, these features are typically available only to specific customers.
Several attempts have been made in the literature to design clock freeze detection sensors in the form of soft IPs for FPGAs.
For instance, Farheen et al.\cite{farheen2022twofold} proposed using internal clock oscillators to monitor the integrity of external clocks. 
More recently, Dumitru et al.\cite{dumitru2025borrowed} introduced two sensor variants that detect clock freezing without relying on any internal clock sources.
 
\parhead{Voltage Sensors.}
Similar to clock sensors, many secure ICs are equipped with voltage sensors to detect voltage tampering~\cite{microchip_security}. 
On FPGAs, analog-to-digital converters (ADCs), such as the XADC available in AMD/Xilinx FPGAs~\cite{xilinx_xadc}, provide built-in voltage sensing capabilities. 
These sensors can monitor both internal and external voltages, converting analog signals into digital values that can be processed by user-defined digital logic on the FPGA.
In addition to built-in voltage sensors, FPGA users can also configure their own delay-based ADCs, such as ring oscillators (ROs)~\cite{zick2012low} and time-to-digital converters (TDCs)~\cite{zick2013sensing}, on the FPGA. 

\parhead{Laser Sensors.}
For specific backscatter-based attacks, such as LLSI, sensors capable of directly detecting incident laser beams have also been investigated. 
Similarly to voltage tampering detection, ADCs have demonstrated sensitivity to localized temperature increases caused by laser irradiation. As a result, ROs~\cite{tajik2017pufmon} and TDCs~\cite{monfared2024laserescape} have been utilized to detect laser probing attacks on FPGAs.

\subsubsection{Response-based Countermeasures}
An often overlooked aspect of countermeasures is the system's response after such powerful attacks are detected.
Although the conventional assumption is that sensitive data can simply be zeroized upon detection, this may not be suitable in many real-world scenarios. 
First, zeroization itself can cause dynamic side-channel leakage~\cite{dumitru2025borrowed}, such as through power side-channels that reveal Hamming weights.
To mitigate this, operations such as masked clear are employed, in which sensitive register contents are overwritten with random values~\cite{dumitru2025borrowed}.
Second, sometimes continued system operation is desired, which could become impossible after zeroization. 
In such cases, schemes such as Moving Target Defense (MTD) can be employed to mitigate the threat without interrupting the system's operation.
For instance, randomizing the placement and routing of a circuit on an FPGA via partial reconfiguration has been shown to be effective against LLSI and IA attacks~\cite{monfared2024laserescape,monfared2024randohm}.

\subsection{Brownout Condition}
When the source voltage of a transistor exceeds a certain threshold, the transistor effectively functions like a switch, responding to changes in the gate voltage. 
Reducing the supply voltage, commonly referred to as voltage scaling, has been widely used to improve the energy efficiency of microprocessors.
The lower bound for voltage scaling is typically constrained to around half of the nominal operating voltage~\cite{haider2008utilizing}.
However, it has been demonstrated that standard CMOS logic gates can operate effectively even below the threshold voltage.
Based on these observations, prototype designs have shown that by carefully replacing analog-like components with standard digital switching elements, it is possible to push voltage scaling into the subthreshold region and extend the traditional limits of low-voltage operation~\cite{sze2006energy,zhai2009energy}.
On commercial FPGAs, however, subthreshold operation is hard to achieve due to the high energy consumption of conventional FPGA interconnects.
While multiple research proposals have been proposed to enhance the performance of subthreshold FPGAs by optimizing interconnect drivers and operating them in the near-threshold voltage region~\cite{pable2011high,qi2016energy,grossmann2012minimum}, they have yet to be realized on commercial FPGAs.

As the supply voltage drops in FPGAs, a brownout condition occurs where the transistors cannot drive the capacitive loads at the gates of other transistors.
Consequently, clock buffers and PLL circuits will also cease to function, and the clock signal distribution will halt.
Meanwhile, memory cells, such as SRAM cells and flip-flops (FFs), may fully retain their data, as their Data Retention Voltage (DRV)~\cite{calhoun2006static,holcomb2012drv} is typically lower than the logic operating threshold. 
For instance, in the case of SRAM-based FPGAs, a brownout condition causes the stoppage of all switching activities within configuration logic blocks (CLBs) and switch boxes. 
At the same time, the SRAMs and flip-flops (FFs) retain the FPGA configuration and user data.
If the voltage drops further, memory cells will lose their content, and thus, the FPGA will crash and require reconfiguration. 
Therefore, a narrow subregion within the subthreshold operating range, above the data retention voltage (DRV), exists where the FPGA enters a ``hibernation" state. 
Furthermore, an FPGA can wake up from hibernation by raising the voltage back above the threshold voltage and resume operation as usual.


\begin{figure}[t]
    \centering
    \includegraphics[width=0.45\textwidth]{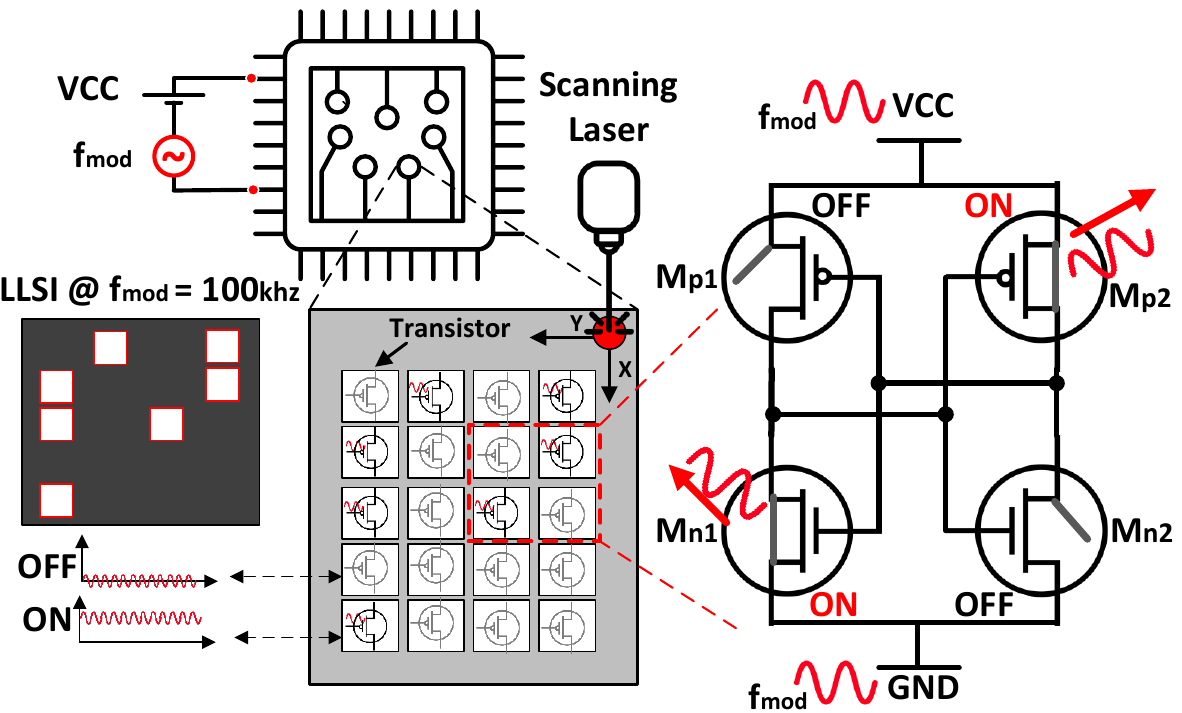}
    \caption{High-level overview of LLSI attack}
    \label{fig:llsi_highlevel}
\end{figure}

\subsection{Laser Logic State Imaging (LLSI)}
Laser Logic State Imaging (LLSI)~\cite{niu2014laser} is a static side-channel attack method.
LLSI is a subset of optical probing attacks, in which a near-infrared laser is focused on the transistors from the backside of an IC, and its reflection becomes modulated by the gate or drain of a transistor during switching activity.
This modulated reflection can be analyzed in two main ways. 
In the first, known as Laser Voltage Probing (LVP), the attacker repeatedly samples the reflection at a single point to reconstruct a waveform of the processed data. 
In the second, known as Laser Voltage Imaging (LVI), the laser is scanned across a region of interest while a spectrum analyzer filters out modulations at a specific frequency.
In a typical LVI setup, the objective is to identify transistors switching at a given frequency and generate a 2D activity map highlighting regions operating at that frequency. 
Combined with LVP, these techniques can reveal internal signal activity, provided that the signals are not static~\cite{tajik2017power,chef2022embedded}.

LLSI builds upon the LVI technique by enabling the probing or imaging of static signals. 
By modulating the power supply at a known frequency, as illustrated in \cref{fig:llsi_highlevel}, the voltage at the transistors that are in ON states will also be modulated, generating a measurable LVI signal. 
In contrast, transistors in the OFF state do not produce a significant signal. As a result, the logic state of a memory cell can be inferred based on the observed LVI activity.
\cref{fig:llsi_highlevel} presents a simplified example of an LLSI image of an SRAM cell composed of two cross-coupled inverters. 
LLSI has been successfully used to extract data from registers in FPGAs~\cite{krachenfels2021real,krachenfels2021automatic} and SRAM cells in microcontrollers~\cite{kiyan2018comparative,chef2021quantitative}.

\subsection{Impedance Analysis (IA)}
Impedance Analysis (IA)~\cite{monfared2023leakyohm} is another static side-channel attack that can recover secret data by measuring changes in the impedance of an IC’s power delivery network (PDN). 
The key idea is that the temporary contents of registers and their corresponding wiring influence the IC's PDN impedance, which leads to changes in how electrical signals with various frequencies reflect or transmit through the IC's PDN. 
To measure such reflections and transmissions, the attacker uses a Vector Network Analyzer (VNA) to inject high-frequency sine waves into the IC's power rails and then captures the so-called scattering parameters. 
Different regions of an IC respond differently at various frequencies~\cite{mosavirik2023silicon}, and thus, by sweeping across a range of frequencies, the attacker can essentially probe multiple localized areas of the chip simultaneously without needing physical access to specific wires or locations. 

\begin{figure}[t]
    \centering
    \includegraphics[width=0.45\textwidth]{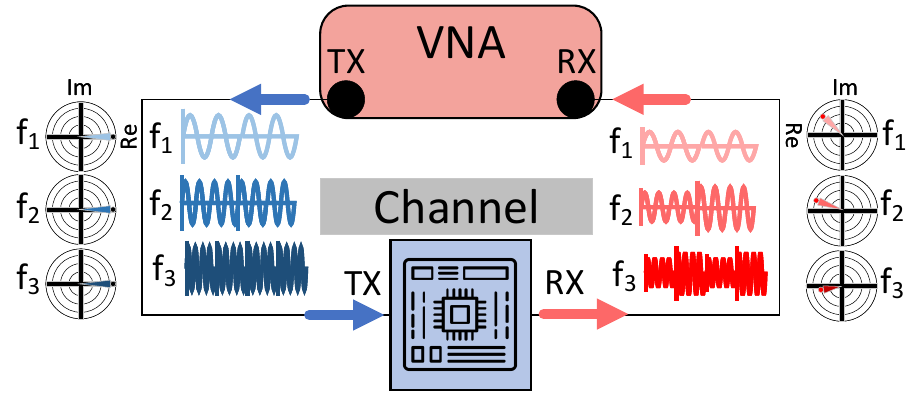}
    \caption{High-level illustration of impedance attack}
    \label{fig:ia}
\end{figure}

\cref{fig:ia} illustrates a high-level overview of an impedance attack. The attack process is conceptually similar to channel estimation in wireless communications, where reference radio frequency (RF) signals are transmitted through a channel and the received signals are analyzed. In the case of impedance attacks, the ``channel" corresponds to the PDN of the target system, while the transmitter and receiver are the ports of a VNA.
RF waves are injected into the PDN at specific frequencies and amplitudes, and the responses are captured at the receiver with amplitude and phase modulations introduced by the circuit's internal state. 
By analyzing these modulated parameters, an attacker can extract secret information. 
Prior work has demonstrated the effectiveness of this approach for attacking both protected and unprotected cryptographic implementations~\cite{monfared2023leakyohm}, as well as for reverse engineering purposes~\cite{awal2023impedance}.



\section{Threat Model}
In our threat model, we assume that the adversary has physical access to the target device. We consider profiling (template) attack scenarios in which the adversary can profile a training device before launching an attack on the actual target. During the attack phase, we assume that all detection- and response-based countermeasures described in \cref{sec:conv_counter} are active.
{We further assume that the security sensors' internals may be unknown to the attacker.}
The adversary does not have access to the system's clock source; however, to read the contents of registers at a specific clock cycle, she must halt the clock.
{{We also assume the adversary can access and manipulate the IC's core voltage supply rails and remove the decoupling capacitors on the printed circuit board (PCB). Such tampering attempts require some knowledge of the PCB schematic, which can be obtained through documentation, visual analysis, or multimeter testing.}}
Moreover, the adversary can capture snapshots of the hardware state using techniques such as LLSI or IA and subsequently recover the values stored in registers.
{{For IA, the attacker must have access to VNAs and function generators, which are standard RF instruments.
For LLSI, failure-analysis equipment can typically be rented for a few hundred dollars per hour rather than purchased outright.
For secret extraction, we consider a template attack threat model: (i) LLSI requires localization of target registers, achievable through reverse-engineering or prior design knowledge; (ii) IA requires knowledge of the cryptographic algorithm, masking scheme, and key size.}}
We also assume the adversary has some knowledge of the system's clock frequency to estimate the brownout voltage thresholds.

To understand how an adversary might benefit from such an attack in the real world, we can consider the following examples.
One example would be the decryption core on FPGAs or microcontrollers/microprocessors, which is programmed with a cryptographic key.
Such decryption engines can be used, for example, to decrypt the device's bitstream or firmware.
By extracting the secret key, the adversary can clone, reverse-engineer, and tamper with the design contained in the bitstream or firmware. 
Moreover, if the same key is used for multiple ICs in the field, the attacker can compromise the security of other ICs that use the same key.

\section{Experimental Setup}
\label{sec:setup}

\subsection{Devices under Test (DUT)}
To test undervolting effects on the clock and voltage sensors, we used two chip families: AMD/Xilinx 7 Series FPGAs and Microchip PolarFire SoC FPGAs, both manufactured using 28 nm technology. While the latter has dedicated hard IP (ASIC) clock and voltage sensors~\cite{microchip_security}, the former has only a hard IP voltage sensor~\cite{peterson2017developing}, and the clock sensor should be configured as a soft IP (e.g., ~\cite{dumitru2025borrowed}).

\subsubsection{AMD/Xilinx FPGAs}
We used NewAE CW305 boards~\cite{cw305_doc}, which have an AMD Artix-7 FPGA (part number XC7A100T-FTG256).
We also used a ChipWhisperer CW310 Bergen Board~\cite{cw310_doc}, which has an AMD Kintex-7 flip-chip FPGA (part number XC7K410T-FBG676).
We selected these boards primarily because they both provide direct access to the FPGA's PDN. 
Furthermore, the OpenTitan design is compatible with the CW310 Bergen Board.
We focused on the $V_\mathit{CCINT}$ power rail, as it directly powers the FPGA's core logic and registers.
Moreover, these boards do not contain any decoupling capacitors on the core voltage PDN, making a rapid voltage drop feasible.

\subsubsection{Microchip FPGA SoCs}
We used a Microchip PolarFire Discovery Kit board, which contains a Microchip PolarFire SoC FPGA (part number MPFS095T-1FCSG325E) \cite{microchip_polarfire}.
We focused on the $V_\mathit{DD}$ power rail, as it directly powers the SoC FPGA's core logic and registers.
For a faster voltage drop, we removed decoupling capacitors C152, C162, C161, C144, C140, C134, and C169, which together account for 491.1 $\mu$F, leaving 0.5308~$\mu$F of decoupling capacitance on $V_\mathit{DD}$.
To gain control over $V_{DD}$, we removed inductor L10 and supplied $V_\mathit{DD}$ through the test point adjacent to the inductor.
We soldered a BNC to a jumper lead cable to the $V_\mathit{DD}$ test point (+) and the ground pad of C144.



\subsection{Optical Setup}


To perform optical probing and photon emission analysis, we used a Hamamatsu PHEMOS-X FA microscope~\cite{hamamatsu_phemosx}. The system is equipped with a 1.3\,{\textmu}m High-power Incoherent Light source (HIL) for optical probing. It supports objective lenses with 5x/0.14\,NA, 20x/0.4\,NA, and 50/0.76\,NA magnifications, along with additional scanner zoom up to 8x.

In LVI/LLSI mode, the laser is scanned across the device's surface using galvanometric mirrors. The reflected light is separated via semi-transparent mirrors and directed to a photodetector. Its output is passed through a bandpass filter tuned to the frequency of interest. The measured signal amplitude at each scan location is mapped to its corresponding spatial position, forming a grayscale-encoded 2D image.

Photon emission is another backside failure analysis technique that captures weak light emissions from the chip's transistors with long exposure times (typically 5\,s or more). 
When current flows through a P-N junction, it can emit a small number of photons due to energy level transitions. For this experiment, we used the PHEMOS-X InGaAs camera operating at –70$^\circ$C with a 20x lens to observe photon emission from a 15-stage ring oscillator implemented on the CW310. This technique was employed to monitor dynamic internal signals, which we expect to become static upon entering hibernation.

\subsection{Side-channel Attack Setups}
We used the CW305 and CW310 boards for all side-channel attacks. \cref{fig:Ia_llsi_setup} depicts the high-level diagram of our experimental setup for IA and LLSI, where the measurements are carried out with an external controller.

\subsubsection{LLSI Attack Setup}




For LLSI, we used the same high-level procedure as shown in \cref{fig:Ia_llsi_setup}. We desoldered the bridge between TP20 and TP19 on the CW310, which cuts off $V_\mathit{CCINT}$ from the onboard voltage regulator. In its place, we connect one channel of a BK Precision 9130 power supply to the SMA connector at J3 (VCCINT\_SHUNTLO). This is because the onboard voltage regulator cannot supply a DC voltage that is low enough to hibernate the FPGA.

The AC modulation comes from a Tektronix AFG3021 single-channel function generator, capacitively coupled through a 10\,$\mu$F electrolytic capacitor soldered to J23 pin~2 on the same side of the shunt as the DC supply. We use a 100\,kHz sinusoidal modulation with an amplitude of roughly 25\,\mV, as measured where the capacitor is connected to $V_{\mathit{CCINT}}$. 
Through trial and error, we found that to be the highest amplitude that would not crash the DUT by bringing the FPGA out of hibernation and into cutoff. 
We achieve this amplitude at the input by setting the function generator output to 1\,\Vpp to compensate for impedance mismatch. 

We programmed the DUT with registers clocked at 10 MHz with an on-chip MMCM.
We can independently set each register to a constant 1, a constant 0, or flip with every clock cycle via a USB serial port through an Arduino (i.e., the controller) connected to the PMOD connectors. 


\subsubsection{IA Attack Setup}
\begin{figure}[t]
    \centering
    \includegraphics[width=0.45\textwidth]{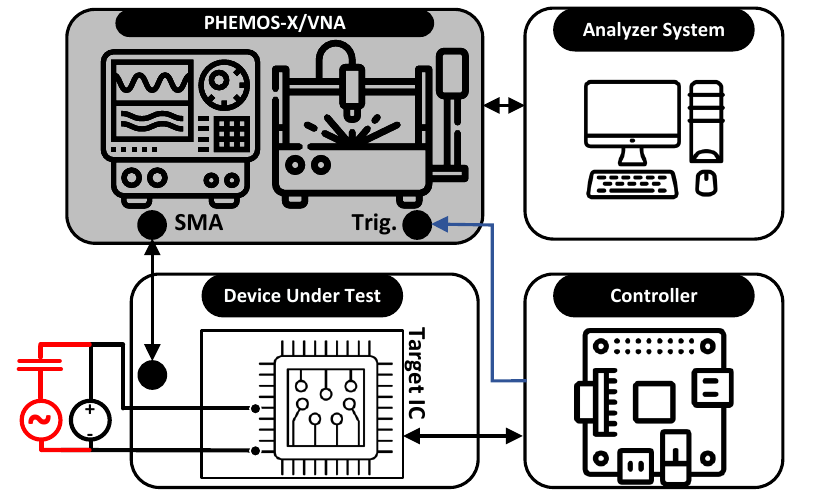}
    \caption{LLSI and IA Setup. Blue parts are used only in IA, red only in LLSI, and the rest are common to both.}
    \label{fig:Ia_llsi_setup}
\end{figure}

To control the state of the target FPGA on the CW305 board during IA, we used a NewAE CW-Lite board~\cite{cw_lite}. It facilitates serial communication with the DUT and acts as an intermediate controller for transferring plaintext and receiving ciphertext from the target IC during profiling. 
The $V_\mathit{CCINT}$ voltage on the CW305 can be controlled through USB and an onboard programmable voltage regulator using Python APIs.
We used a Keysight ENA Network Analyzer E5080A~\cite{Keysight_VNA} for our measurements, which supports RF/microwave scattering analysis across frequencies ranging from 9 kHz to 6 GHz. 
To ensure reliable signal transmission, we used Minicircuit CBL-2FT-SMNM+ shielded characterization cables~\cite{shielded_cable}.

After loading a bitstream into the target FPGA, arbitrary input data (e.g., plaintexts and keys) are generated by the Analyzer System and sent to the Controller, as shown in \cref{fig:Ia_llsi_setup}. 
The controller transmits this data to the target IC via a serial interface. At a designated timestamp, the system drops the voltage to pause the clock; then, the controller triggers the VNA to capture a measurement. The VNA then collects the trace and returns it to the Analyzer System. 

The VNA measurement parameters in our analysis are determined experimentally. The intermediate frequency (IF) bandwidth is set to 500\,Hz, and the averaging factor is configured as $N_{\text{Avg}} = 400$ to reduce the measurement noise. 
We configure the VNA to perform a single-port measurement of $S_{11}$, the reflection scattering parameter, which provides a linear estimation of the impedance profile. 
Although both amplitude and phase values of the scattering parameters are recorded, we use only the phase component in our impedance analysis due to its noise resiliency~\cite{mosavirik2023silicon}. 
All measurements are performed differentially using a reference program, and the VNA output power is set to 10\,dBm.
Upon completion of the computation on the target IC, the controller receives the ciphertexts and forwards them back to the Analyzer System over a serial connection for verification. 

\section{Circuit Operation During Hibernation}



\subsection{Hibernation Voltage Characterization}
\label{sec:characterization}

\subsubsection{Undervolting Voltage-Frequency Scan}
To characterize the resilience of FPGA targets under voltage stress, we developed a systematic \textit{Hibernation Scan} methodology that explores the functional limits of an FPGA's internal logic as supply voltage decreases across a range of clock frequencies. This method reveals the threshold conditions under which the FPGA ceases to reliably perform core operations such as register assignment and clock-driven state transitions. 
The procedure is executed through a software controller that communicates with an on-chip UART-based hardware test module implemented in RTL within the target FPGA DUT.
Crucially, this UART module is on a separate voltage rail to the FPGA internal logic, so it remains functional during our undervolting experiments.

\parhead{Hardware Instrumentation.}
The FPGA on the CW305 board is configured with a test logic block that performs two key functions during a timed evaluation window of undervolting. First, a known \texttt{register assignment}, e.g., \texttt{reg\_out\,<=\,8'h88}, that allows for deterministic verification of data latching under voltage stress is executed. Then, a \texttt{clock counter} that increments on every rising clock edge, validates whether the internal clock network and FFs remain operational.

\parhead{Initialization Procedure.}
On the host side, the undervolting scan is orchestrated via a high-level script. The scan proceeds by sweeping a grid of \texttt{(frequency, voltage)} pairs.
The PLL is configured to the desired frequency $f$ at the supply voltage $v$.


Furthermore, we deploy a Debug Register Reset routine at RTL level that resets all registers to a known baseline value. 
Before each iteration of the sweep, two time parameters are set. $t_d$ as the FPGA initialization delay before evaluation (the delay incurred due to UART communication for send commands), and $t_t$ as the duration in which the FPGA is undervolted.


\parhead{Undervolting Phase.} When the test is initialized, the supply voltage is simultaneously lowered to $v$ via the programmable on-board power supply. The chip is allowed to run in the undervolted state for a minimum of $t=0.8$ seconds, ensuring that the on-chip execution occurs during the undervolted state ($t > t_d + t_t$). After the undervolting phase, the voltage is restored to a nominal level (typically 1.0\,V), and the contents of the test registers are read back. The register assigned value (\texttt{reg\_assign}) and the clock counter state (\texttt{clock\_count}) from FPGA are extracted via UART. Moreover, we verify system stability by performing the hardware sanity test. If the FPGA does not return to a clean baseline state, it is assumed to have entered a permanent fault state (crashed). In this case we reprogram the FPGA bitstream to recover from the crash. For each \texttt{(f, v)} pair, the results are logged, including the observed values of \texttt{reg\_assign}, \texttt{clock\_count}, and a boolean \texttt{crash} flag indicating system failure. The details of the undervolting scan is described in \cref{apx:alg1}.

\begin{figure}[t]
    \centering
    \includegraphics[width=0.45\textwidth]{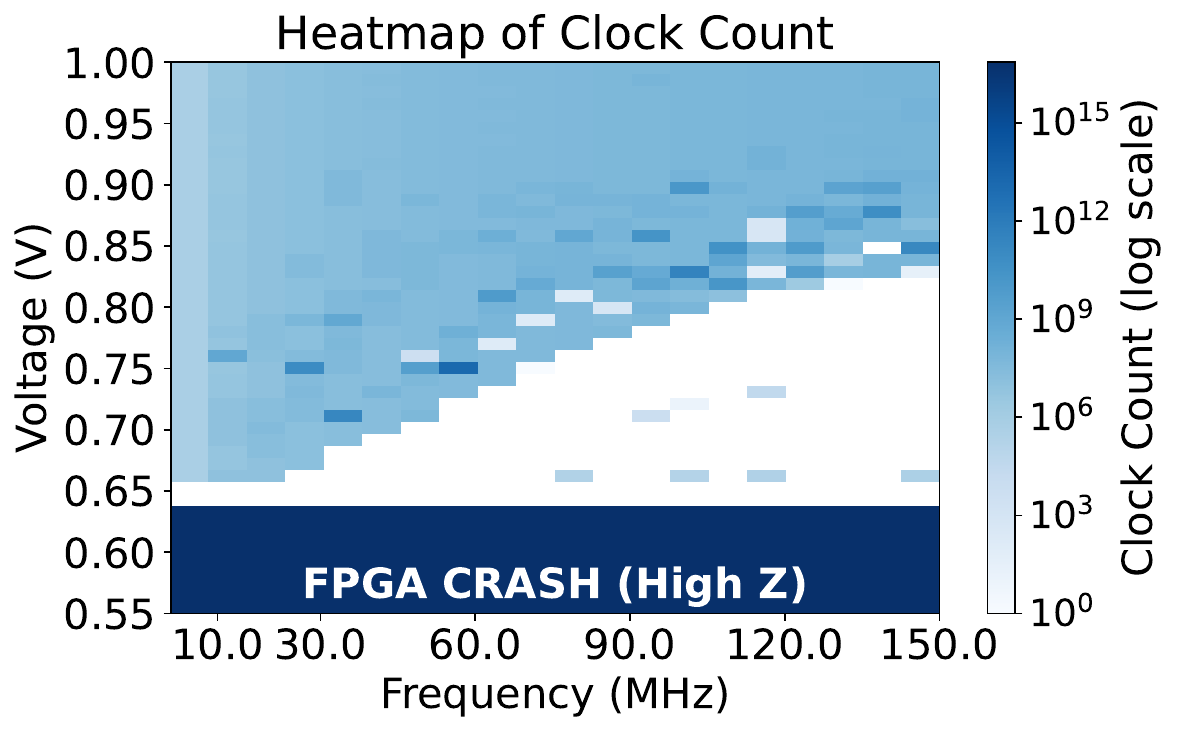}
    \caption{Heatmap of the \texttt{clock\_count} during undervaluing with different working clock frequencies. White spots highlight the hibernation voltage.  }
    \label{fig:clock-heatmap}
\end{figure}

\begin{figure}[t]
    \centering
    \includegraphics[width=0.45\textwidth]{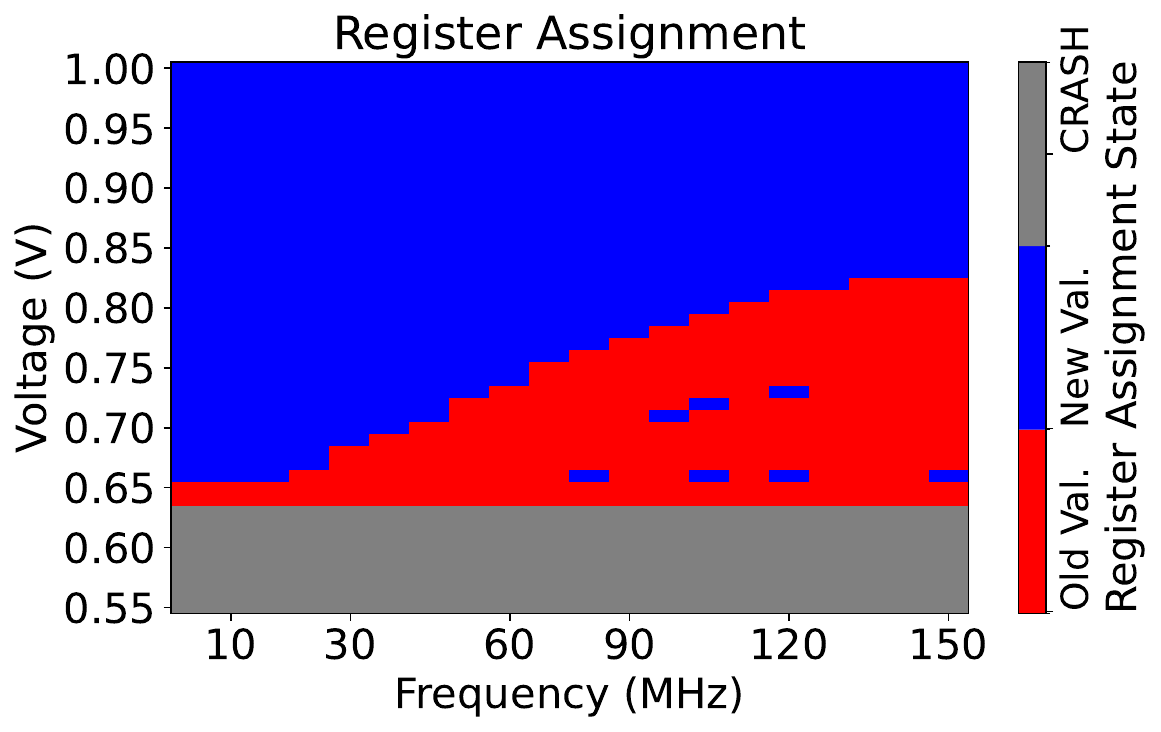}
    \caption{Heatmap of the \texttt{reg\_assign} during undervaluing with different working clock frequencies.}
    \label{fig:register-heatmap}
\end{figure}





\subsubsection{Clock Count Reliability Under Undervolting}

\cref{fig:clock-heatmap} illustrates the behavior of the internal clock counter across a sweep of operating frequencies and supply voltages. The heatmap encodes the \texttt{clock\_count} output using a logarithmic scale, where darker shades represent higher accumulated counts during the evaluation phase, and lighter shades indicate degraded or failed clock operation. White regions correspond to clock counts below the minimum detectable threshold and are considered \textit{Hibernated} states.

At nominal voltage levels (above 0.85\,V), the clock operates reliably across the entire frequency range up to 150\,MHz, as evidenced by the uniform high count values (light blue regions). As the voltage decreases, however, a clear degradation pattern emerges: higher frequencies are the first to experience failure, while lower frequencies maintain functionality down to lower voltage thresholds.
The transition boundary from dark to light regions represents the onset of clock instability and is referred to as the \textit{hibernation voltage}. This is the minimum voltage at which the clock counter can still increment meaningfully at a given frequency. Below this boundary, the white regions indicate complete clock failure—either due to the PLL losing lock, internal flip-flops ceasing to toggle, or propagation delays exceeding the clock period.
The stable dark blue strip at the bottom of the plot (around 0.63\,V and below) is a result of a complete FPGA crash where it cannot be recovered and captured data are all high-impedance (i.e., \texttt{0xff}).

\subsubsection{Register Assignment Reliability}

\cref{fig:register-heatmap} presents a discrete heatmap characterizing the behavior of the FPGA’s register assignment under varying voltage and frequency conditions. Each cell represents the observed value of a target register after the undervolting test phase, with three possible states: a correct new assignment (blue), an old or default value indicating a failure to assign (red), or a system crash (gray).

At voltages above 0.85\,V, the register reliably latches the expected value (\texttt{0x88}) across all operating frequencies, indicating stable sequential logic and reliable data propagation. However, as the voltage drops below 0.70\,V, an increasing number of cells transition to red, particularly at higher frequencies. This transition indicates a failure in the FPGA’s ability to commit new values to registers—likely due to setup/hold time violations, degraded signal swing, or metastability induced by reduced supply voltage.
Below approximately 0.60\,V, the majority of register operations either result in incorrect values or trigger system crashes. These regions highlight the lower bound of operational safety for secure register logic.

The security implications of such failures are significant. Many FPGA-based protection mechanisms, including cryptographic randomization, register obfuscation, or randomized key preloading, rely on the ability to overwrite internal state deterministically.
If undervolting prevents register assignments from executing correctly, it opens the possibility of residual sensitive data being left behind and making the system vulnerable to static SCA such as IA and LLSI.


\begin{figure}[t]
    \setlength{\abovecaptionskip}{0pt}
    \setlength{\belowcaptionskip}{0pt} 
    
    \centering
    \begin{subfigure}{0.238\textwidth}
       \setlength{\abovecaptionskip}{0pt}
        \centering
        \includegraphics[width=\textwidth]{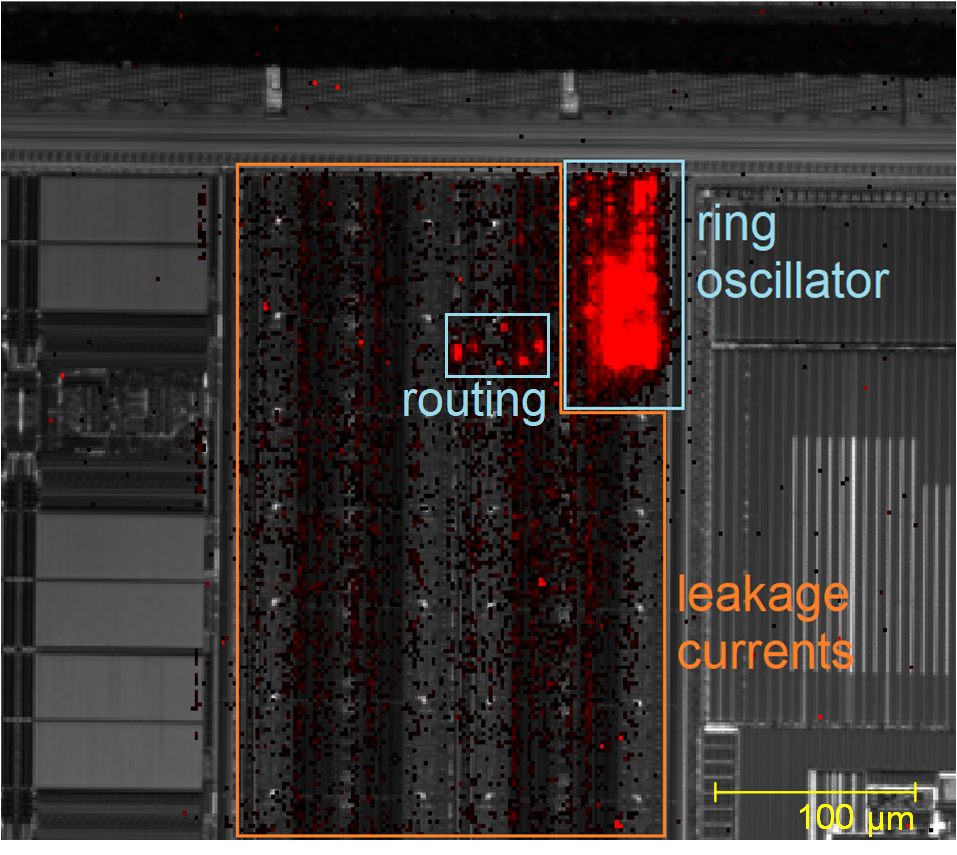}
        \caption{}
        \label{fig:ro_emi:on}
    \end{subfigure}
    \hfill
    \begin{subfigure}{0.238\textwidth}
       \setlength{\abovecaptionskip}{0pt}
        \centering
        \includegraphics[width=\textwidth]{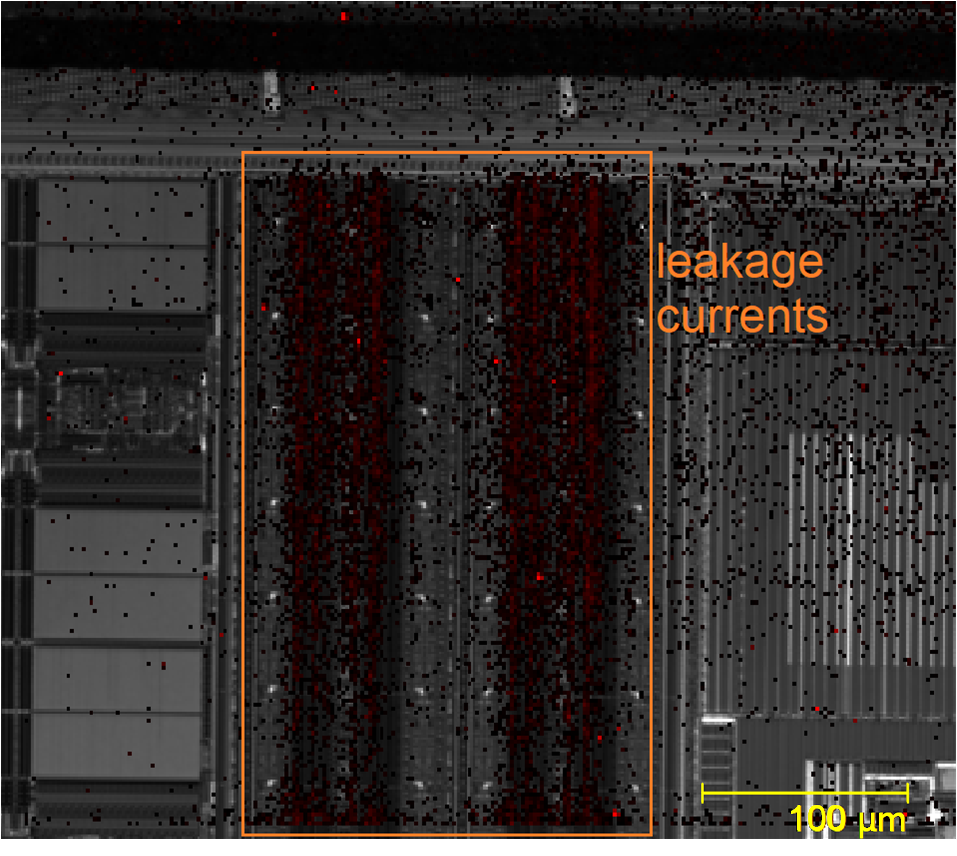}
        \caption{}
        \label{fig:ro_emi:off}
    \end{subfigure}
    \caption{Photon emission images of a ring oscillator on a Kintex-7 at (a) nominal ($V_{\mathit{CCINT}}$=1000\,\mV) and (b) hibernation ($V_{\mathit{CCINT}}$=555\,\mV) voltages}
    \label{fig:ro_emi}
\end{figure}

\subsection{Verifying Hibernation with Photon Emission}\label{sec:clock_PE}
Relying on the FPGA IOs to verify that circuit switching is disabled may not be reliable, as an undervolted DUT may not be able to drive the IO buffers due to them being a large capacitive load. 
Hence, we can perform photon emission analysis to measure the activity to verify disabled circuits without relying on the chip itself.
Using photon emission, we can observe dynamic internal signals, which should become static when entering hibernation.


As a test circuit, we chose a ring oscillator because it has a high emission rate due to its fast switching gates, and it is also the building block of various clock and voltage sensors~\cite{farheen2022twofold}.
We can see the photon emission of the ring oscillator, as shown in \cref{fig:ro_emi:on}. 
After lowering $V_\mathit{CCINT}$ to hibernation levels, we find that the ring oscillator disappears, leaving only the leakage currents shown in \cref{fig:ro_emi:off}. 
The presence of leakage currents during hibernation suggests that photon emission still occurs under these conditions due to the retained FPGA configuration.
Therefore, the disappearance of the ring oscillator in the emission images suggests that the undervolting operation entirely disables its switching activity.

\subsection{LVI of Clock Buffers}
\begin{figure}[t]
    \setlength{\abovecaptionskip}{0pt}
    \setlength{\belowcaptionskip}{0pt} 
    
    \centering
    \begin{subfigure}{0.22\textwidth}
       \setlength{\abovecaptionskip}{0pt}
        \centering
        \includegraphics[width=\textwidth]{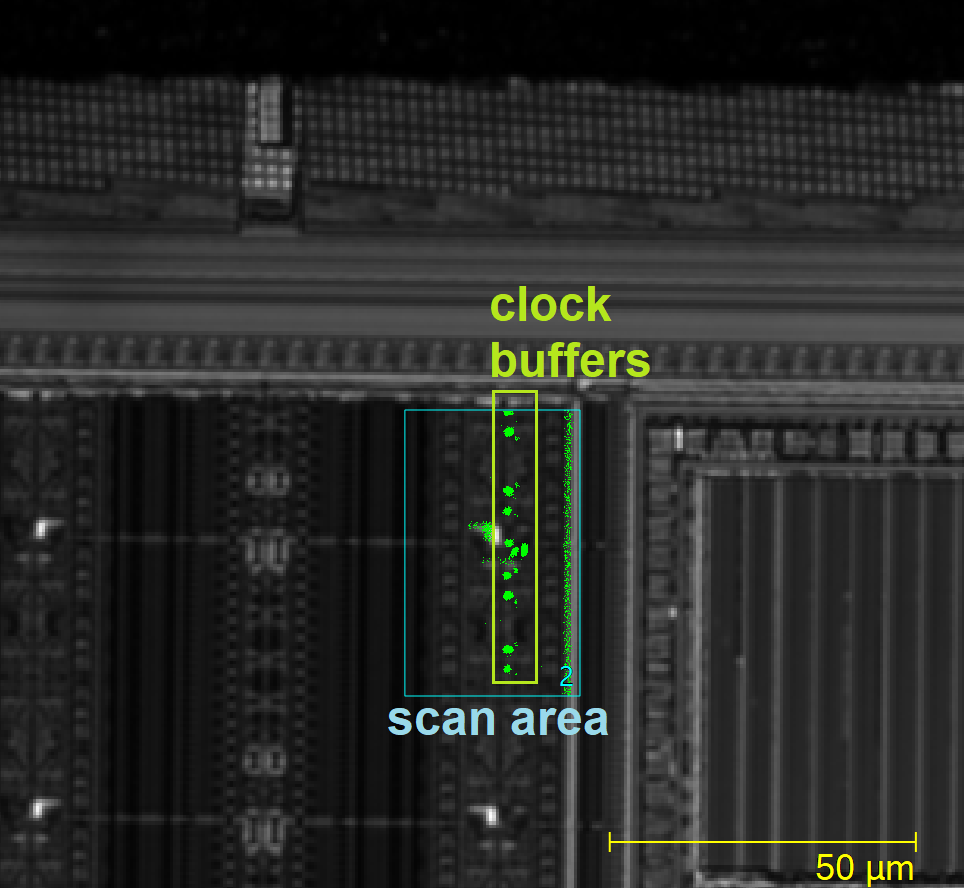}
        \captionsetup{font=footnotesize}
        \caption{}
        \label{fig:clk_lvi:on}
    \end{subfigure}
    \hfill
    \begin{subfigure}{0.22\textwidth}
       \setlength{\abovecaptionskip}{0pt}
        \centering
        \includegraphics[width=\textwidth]{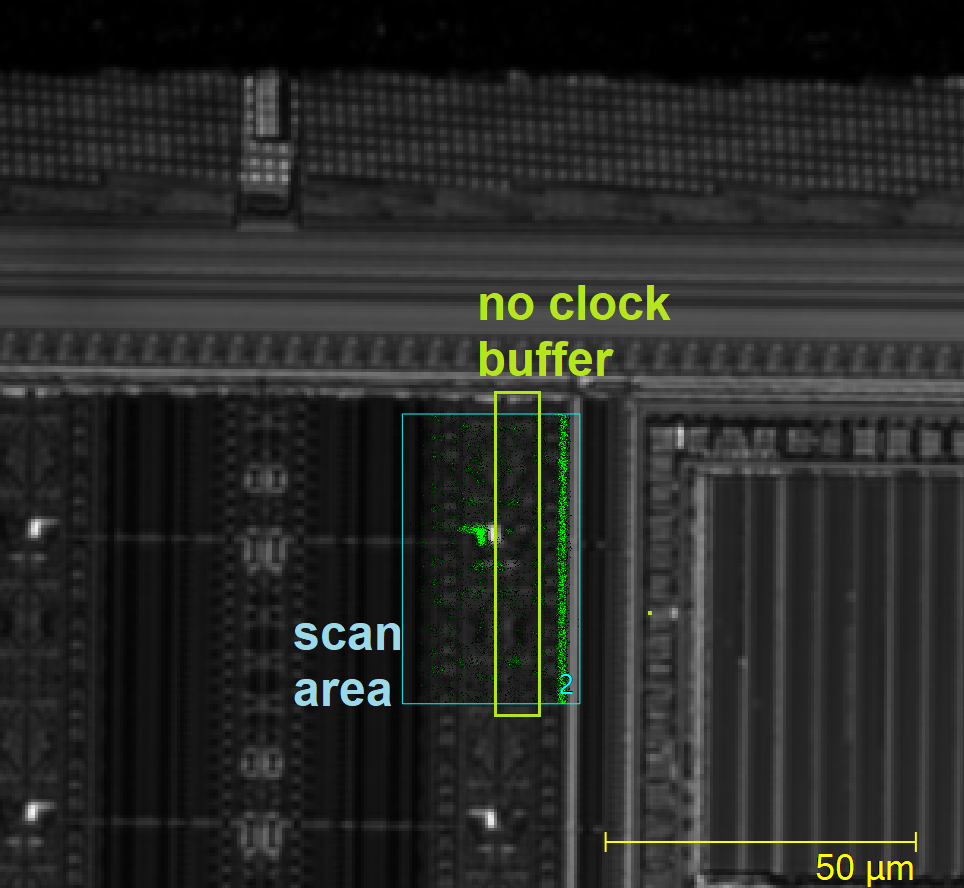}
        \captionsetup{font=footnotesize}
        \caption{}
        \label{fig:clk_lvi:off}
    \end{subfigure}
    \caption{LVI images of some Kintex-7 clock buffers at (a) nominal ($V_{\mathit{CCINT}}$=1000\,\mV) and (b) hibernation $V_{\mathit{CCINT}}$=555\,\mV) voltages}
    \label{fig:ro_emi}
\end{figure}

Another method to verify that internal clocks are disabled during hibernation is through LVI. We implemented four registers on the CW310, clocked at 10\,MHz by an on-chip MMCM. Using the high-power incoherent light HIL and a 50× objective lens on the PHEMOS-X microscope, we performed an LVI scan at 10\,MHz to locate the clock buffers. Under normal operating conditions, we observed several bright spots corresponding to active clock buffers, as shown in \cref{fig:clk_lvi:on}.
However, when $V_\mathit{CCINT}$ is lowered to hibernation levels, these bright spots disappear, as illustrated in \cref{fig:clk_lvi:off}. This disappearance indicates that the clock buffers are no longer switching at 10\,MHz, confirming that the registers are effectively disabled. This effect may be attributed to the MMCM either losing its ability to drive the clock network or being entirely disabled under hibernation.

\subsection{Verifying Hibernation of Flash-Based FPGAs}

\begin{figure*}[t]
\captionsetup[subfigure]{labelformat=empty}
    \centering
    \setlength{\abovecaptionskip}{0pt}
    
    \begin{subfigure}[b]{0.4\textwidth}
        \includegraphics[width=\linewidth,keepaspectratio]{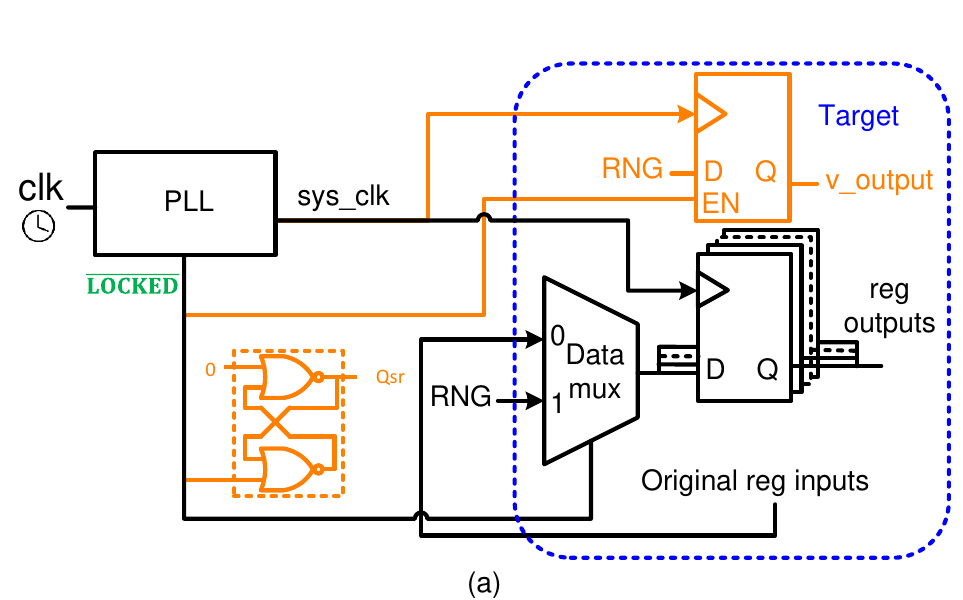}
        \caption{}
        \label{fig:BT_PLL}
    \end{subfigure}
    \hfill
     \begin{subfigure}[b]{0.58\textwidth}
     \includegraphics[width=\linewidth,keepaspectratio]{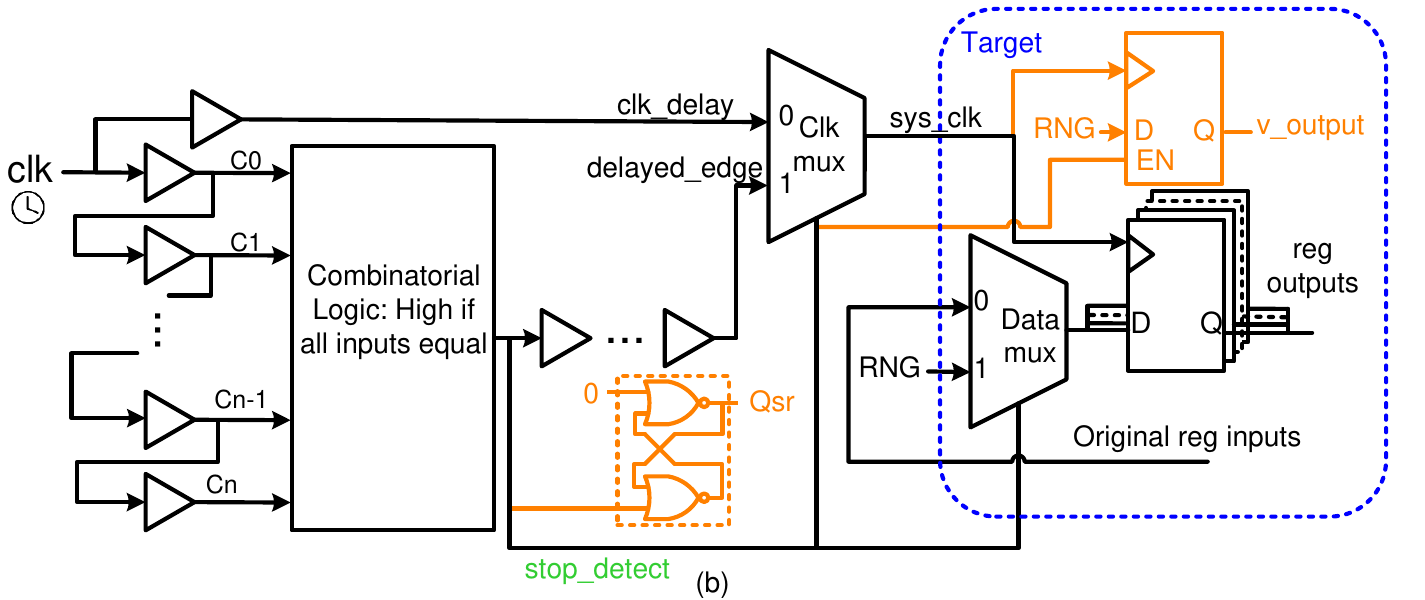}
        \caption{}
        \label{fig:BT_Asynchronous}
    \end{subfigure}

    \caption{Borrowed Time countermeasures: (a) PLL-based clock sensor, (b) Delay chain-based sensor. The SR latch and the register (in orange) detect if the alarm goes high and if register latching occurs in an undervolting-induced clock halt.}
    \label{fig:BT}
\end{figure*}

Although the flash-based PolarFire's configuration is non-volatile, unlike the SRAM-based FPGAs, the data stored in its registers is volatile. To characterize the hibernation range of the PolarFire, we run a counter, manually lower $V_{DD}$, and then restore it to its nominal value. If the FPGA does not crash, the counter value should remain unchanged regardless of the duration of the hibernation, and the device should resume operation from that exact point once power is restored. Based on these experiments, we observe that the PolarFire SoC FPGA supports hibernation across a wide voltage range, approximately from 0.9\,V down to 0.2\,V.

\section{Defeating Sensors}\label{sec:defeat}



\subsection{Defeating Clock Sensors (Soft IPs)}
\label{sec:BTdefeat}

Devices run on clocks that are either externally sourced, such as from crystal oscillators, or generated internally by oscillator circuitry. 
Several countermeasures based on clock sensors have been designed to protect systems from clock tampering.
This is vital as clocks can be a single point of failure for security mechanisms that depend on reliable state modification like cryptographic randomization, obfuscated registers, and randomized key loading. 
Data retention under stopped clock conditions also exposes sensitive values to static side-channel analysis. 
Here, we focus on a recently proposed countermeasure~\cite{dumitru2025borrowed} specifically designed to mitigate this threat. 
The function of this countermeasure is two-fold: it monitors the system clock such that it can trigger an alarm upon detecting a stop condition, then, in response to this alarm, it performs some actions to clear sensitive data from the circuit. 
This has only been evaluated on target circuits operating with a nominal supply voltage. 
We investigate its ability to perform both its detection and clearing functions when undervolted by Chypnosis. 
We briefly explain both variants of the countermeasure presented in~\cite{dumitru2025borrowed}, to contextualize our findings. 

\subsubsection{PLL-based Clock Sensor}
The first variant uses a Phase-Locked Loop (PLL), see \cref{fig:BT_PLL}.
This is a standard clock management control circuit used to generate clock signals derived from an externally sourced reference clock. 
PLLs will typically include an output status signal to indicate whether their generated clocks are synchronized, or ``locked", to the reference. 
A stopped input clock will be detected rapidly by a PLL and signaled via deassertion of this signal, so~\cite{dumitru2025borrowed} proposes using this as an alarm. 

To evaluate the PLL-based sensor, we port the openly available implementation~\cite{BorrowedTime_github} to our CW310 board. 
This is designed for 7-Series AMD/Xilinx FPGAs and involves a Mixed-Mode Clock Manager (MMCM) with a \texttt{LOCKED} status signal from its internal PLL. 
The system then uses an alarm based on the inverse of this signal, $\overline{\text{\texttt{LOCKED}}}$, to switch the input of sensitive registers to instead come from an RNG. 
When input has been switched, subsequently arriving clock pulses from the PLL output latch the random values to clear sensitive data, this is called a \emph{masked clear}.
We first confirm the detection and clearing of our ported implementation to work correctly at the nominal supply voltage.

From preliminary Chypnosis experiments with a 10\,MHz clock and setting $V_{\mathit{CCINT}}$ to a 0.555\,V hibernation voltage, we find that the protection mechanism fails. 
For system introspection to find the cause, we add an SR latch that takes the alarm as input, and a register that takes the alarm as a latch enable signal, both are shown in orange in \cref{fig:BT_PLL}. 
When we perform Chypnosis again, we find the SR latched value to be high, indicating successful detection.
We also find the register output (\texttt{V\_output}) retained its default value, suggesting that the mechanism to latch random data into the register failed.
We further confirm the clock had indeed stopped using photo emission and laser probing of \texttt{sys\_clk} as in \cref{sec:clock_PE}.
These results demonstrate that while the PLL-based protection can detect a stopped clock, it can still be bypassed by undervolting due to its failure to perform the masked clear (data overwrite) response actions.

\subsubsection{Asynchronous Clock Sensor}

For systems where PLLs are unsuitable, such as in low-power designs or systems with clock gating, \cite{dumitru2025borrowed} proposes a different clock sensor system based on signal propagation delay, see \cref{fig:BT_Asynchronous}.
The clock sensor is based on a tapped delay chain that takes the clock as the input, and effectively samples it at various time offsets defined by the delays between taps. 
All of these taps are fed into a combinatorial element that asserts an alarm signal if all inputs are equal, which is the case if the clock is stopped. 
The system works similarly to the PLL solution in that it uses the alarm, \texttt{stop\_detect}, to select randomness as the input to sensitive registers.
However, this design does not have a transitioning clock source to use for latching the randomness, thus the system also needs circuitry for generating an active clock edge.
This is based on a secondary delay chain that takes the alarm signal, since this provides a low-to-high transition upon an alarm. 
The clock source going into the target, \texttt{sys\_clk}, is multiplexed between the original \texttt{clk} and \texttt{delayed\_edge}.

We port the asynchronous delay-based sensor countermeasure variant~\cite{BorrowedTime_github} to our CW310 board, and again confirm that the countermeasure works correctly at nominal supply voltage.
Initial Chypnosis experiments, as carried out against the PLL-based system, similarly show the protection mechanism of this variant to fail. 
We modify the circuit with a similarly placed SR latch and register (shown in orange in \cref{fig:BT_Asynchronous}) for introspection.
Then, when performing Chypnosis again, we observe the same result as for the PLL system, which is that the detection mechanism functions correctly to trigger an alarm, but subsequent latching for the masked clear does not complete. 
We include a timing diagram of internal signals in \cref{sec:timingdiagram}.
We again confirm the clock had stopped using photon emission and laser probing of \texttt{sys\_clk} as in \cref{sec:clock_PE}.
Thereby, we can also bypass this countermeasure variant with undervolting.

\subsection{Defeating Xilinx Voltage Sensor (Hard IP)}\label{sec:xilinx_xadc}

\begin{figure}[t]
    \centering
    \setlength{\abovecaptionskip}{0pt}
    \includegraphics[trim=10 10 10 10, clip, width=0.45\textwidth]{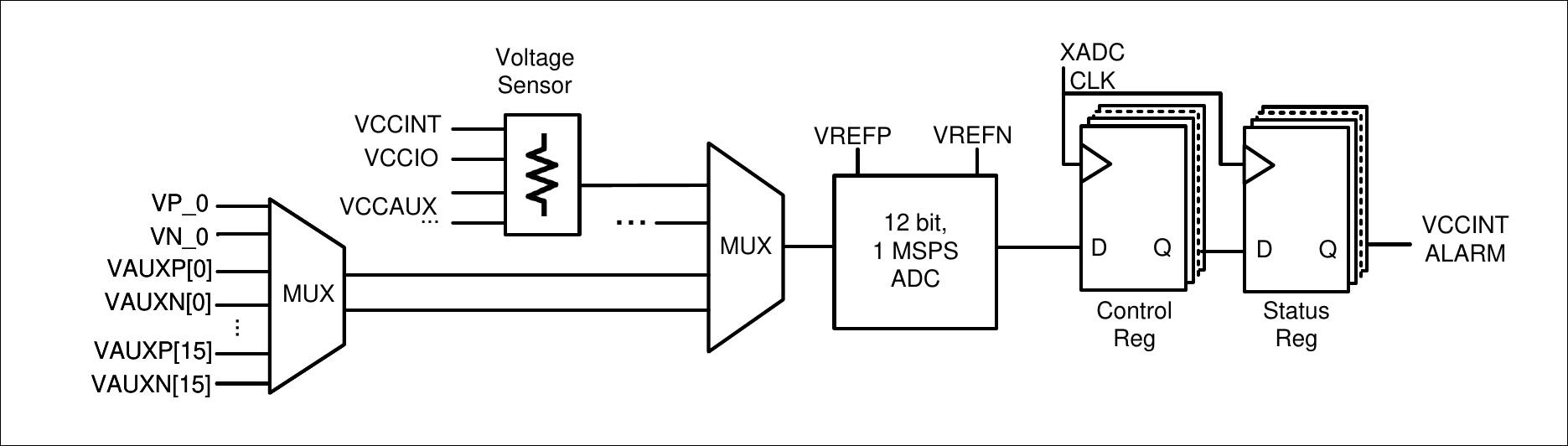}
    \caption{Block diagram of the XADC on 7 Series FPGAs} 
    \label{fig:xadc_setup}
\end{figure}

AMD/Xilinx 7 Series FPGAs are equipped with an on-chip voltage sensor for $V_{\mathit{CCINT}}$, which is connected to the XADC, a 12-bit, 1 Mega Samples Per Second (MSPS) analog-to-digital converter~\cite{xilinx_xadc}. Through a multiplexer, the XADC also shares its functionality with a temperature sensor, other voltage rail sensors (e.g., $V_{\mathit{CCAUX}}$ and $V_{\mathit{CCIO}}$), and various analog I/Os, as illustrated in \cref{fig:xadc_setup}.

\begin{figure}[t]
    \setlength{\abovecaptionskip}{0pt}
    \setlength{\belowcaptionskip}{0pt} 
    
    \centering
    \begin{subfigure}{0.23\textwidth}
        \setlength{\abovecaptionskip}{0pt}
        \centering
        \includegraphics[trim=10 20 20 20, clip, width=\textwidth]{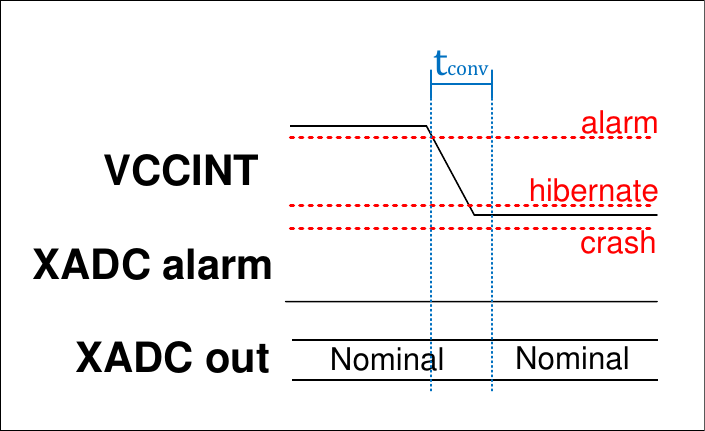}
        \caption{Success}
        \label{fig:xadc_defeat:success}
    \end{subfigure}
    \hfill
    \begin{subfigure}{0.23\textwidth}
        \setlength{\abovecaptionskip}{0pt}
        \centering
        \includegraphics[trim=10 20 20 20, clip, width=\textwidth]{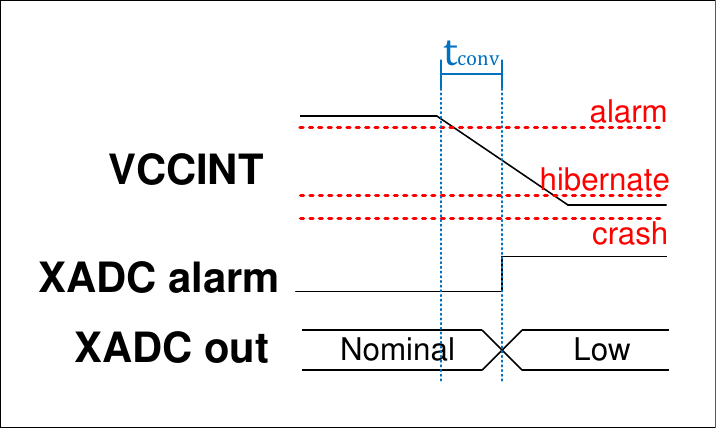}
        \caption{Failure}
        \label{fig:xadc_defeat:fail}
    \end{subfigure}
    \caption{Graph of $VCC_{INT}$ and associated XADC readings for an attempt to defeat the XADC}
    \label{fig:xadc_defeat}
\end{figure}

\begin{figure}[t]
    \setlength{\abovecaptionskip}{0pt}
    \setlength{\belowcaptionskip}{0pt} 
    
    \centering
    \begin{subfigure}{0.237\textwidth}
        \setlength{\abovecaptionskip}{0pt}
        \centering
        \includegraphics[trim=165 340 185 340, clip, width=\textwidth]{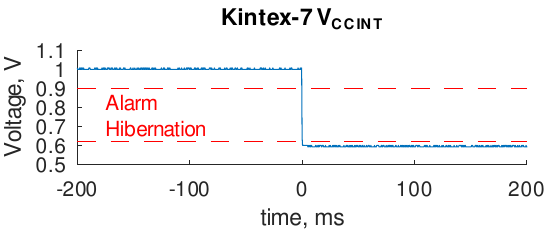}
        \caption{Success}
        \label{fig:xadc_scope:success}
    \end{subfigure}
    \hfill
    \begin{subfigure}{0.237\textwidth}
        \setlength{\abovecaptionskip}{0pt}
        \centering
        \includegraphics[trim=165 340 185 340, clip, width=\textwidth]{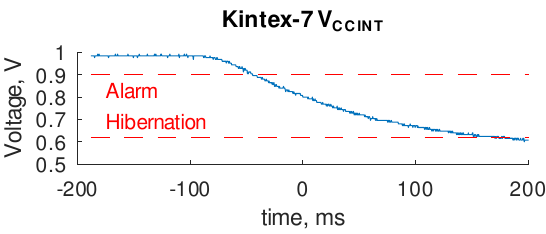}
        \caption{Failure}
        \label{fig:xadc_scope:fail}
    \end{subfigure}
    \caption{Oscilloscope waveforms for $V_{\mathit{CCINT}}$ when attempting to defeat the XADC alarm signal}
    \label{fig:xadc_scope}
\end{figure}

The XADC features a built-in voltage alarm that triggers when the voltage falls outside a user-defined range, from $V_{\mathit{alarmL}}$ to $V_{\mathit{alarmH}}$. However, these alarms rely on the digital output of the XADC. If the $V_{\mathit{CCINT}}$ voltage is dropped rapidly enough, such that the time between $V_{\mathit{HIB}}$ and $V_{\mathit{alarmL}}$ is shorter than the XADC's conversion time ($T_{\mathit{conv}}$), the alarm will not be triggered. If successful, the last XADC reading before hibernation will show the normal operating voltage, as shown in \cref{fig:xadc_defeat:success}.

When using a BK Precision 9130 triple-output power supply to power the ChipWhisperer CW310 through SMA connector J3, we were able to reduce $V_{CCINT}$ to hibernation levels in approximately 400\,ms.
However, this drop was too slow to bypass detection, and the XADC alarm signal was triggered, as shown in \cref{fig:xadc_scope:fail}.
In contrast, by using a Tektronix AFG 3021 function generator as the power source, we achieved a much faster voltage drop to $V_{\mathit{HIB}}$ within 80\,$\mu$s---sufficient to avoid triggering the alarm signal (see \cref{fig:xadc_scope:success}).
Although the XADC's sample rate of 1\,MSPS suggests a conversion time ($T_{\mathit{conv}}$) of 1\,$\mu$s, the actual conversion time may be longer due to pipelining in the ADC architecture~\cite{adi_adc}.
Pipelining allows higher sample rates but introduces additional latency. While the XADC documentation does not explicitly mention its latency or pipelined design, our experimental results suggest that pipelining is employed.
We also note that the XADC is not outright disabled during undervolting, as it is 
powered by $V_{\mathit{CCAUX}}$ (1.8\,V) rather than $V_{\mathit{CCINT}}$.
Instead, it is more likely that the associated control registers become disabled, effectively rendering the XADC non-functional during hibernation.

\subsection{Defeating Microchip AT Module (Hard IP)}

\begin{figure}[t]
    \centering
    \setlength{\abovecaptionskip}{0pt}
    \includegraphics[trim=10 10 10 10, clip, width=0.45\textwidth]{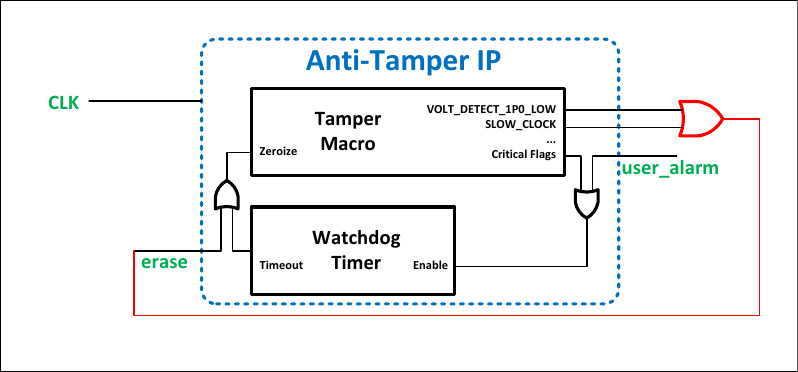}
    \caption{A simplified block diagram of the AT module on Microchip PolarFire SoC FPGAs} 
    \label{fig:polarfire_tampermodule}
\end{figure}

Microchip PolarFire SoC FPGAs are equipped with a dedicated anti-tamper (AT) module~\cite{microchip_polarfire} which consists of, among other things, on-chip voltage and clock sensors. These sensors can then zeroize the device, erasing the programmed bitstream and all data. Unlike the XADC on Xilinx devices, these sensors have fixed low and high voltage alarm thresholds, which assert the \verb|VOLT_DETECT_1P0_LOW| and \verb|VOLT_DETECT_1P0_HIGH| flags, respectively. In addition, the system's controller clock slows from 80 MHz to 20 MHz when it detects a $V_{DD}$ brownout, and the tamper macro asserts the \verb|SLOW_CLOCK| flag, which effectively serves as a second low-voltage alarm signal.
The AT module monitors various tamper flags, which can be set to zeroize the FPGA in one of two ways. There is a watchdog timer that waits 1000 clock cycles after being enabled before zeroizing the device. For more critical tamper events, an additional input is available to zeroize the device immediately. To ensure the fastest possible zeroization, we connect the VOLT$\_$DETECT$\_$1P0$\_$LOW and SLOW$\_$CLOCK flags to the latter input (see \cref{fig:polarfire_tampermodule}).

\begin{figure}[t]
    \setlength{\abovecaptionskip}{0pt}
    \setlength{\belowcaptionskip}{0pt} 
    
    \centering
    \begin{subfigure}{0.237\textwidth}
        \setlength{\abovecaptionskip}{0pt}
        \centering
        \includegraphics[trim=165 340 185 340, clip, width=\textwidth]{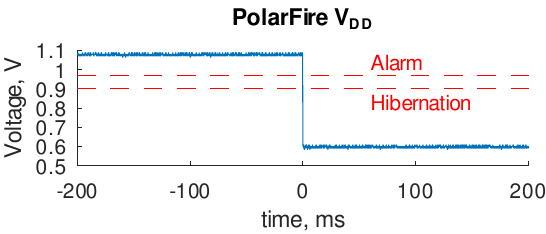}
        \caption{Success}
        \label{fig:mpfstamper_scope:success}
    \end{subfigure}
    \hfill
    \begin{subfigure}{0.237\textwidth}
        \setlength{\abovecaptionskip}{0pt}
        \centering
        \includegraphics[trim=165 340 185 340, clip, width=\textwidth]{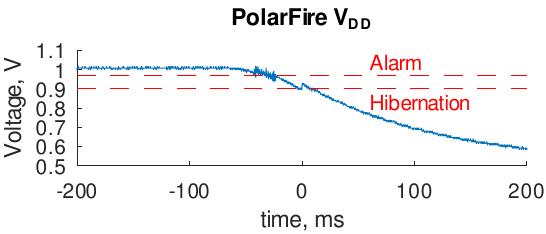}
        \caption{Failure}
        \label{fig:mpfstamper_scope:fail}
    \end{subfigure}
    \caption{Oscilloscope waveforms for $V_{\mathit{DD}}$ when attempting to defeat the PolarFire anti-tamper module}
    \label{fig:mpfstamper_scope}
\end{figure}

Using a BK Precision 9130 triple-output power supply to power the modified Microchip Kit, we were able to reduce $V_{DD}$ to $V_{\mathit{HIB}}$ in approximately 61.6,ms (see \cref{fig:mpfstamper_scope:fail}). However, this voltage drop was too slow to bypass detection, resulting in both the \verb|SLOW_CLOCK| and \verb|VOLT_DETECT_1P0_LOW| flags being triggered. When configured to do so, these flags can initiate zeroization. In contrast, using a Tektronix AFG 3021 function generator as the power source enabled a much faster voltage drop (approximately 430\,ns to reach $V_{\mathit{HIB}}$) which was fast enough to avoid triggering either flag (see \cref{fig:mpfstamper_scope:success}). Interestingly, even when the flags are configured to trigger zeroization, it does not occur in this rapid-drop scenario. Additionally, if we disable zeroization and set these flags to write to a register, no data is written in the rapid-drop scenario.

\subsection{Voltage Falling Time vs Clock Frequency}

{We explore precisely how undervolting fall time influences attack success.
Unfortunately, our equipment and setups do not support customizable voltage falling times. 
However, we recall that we expect rapid undervolting over a short time frame succeeds in bypassing voltage sensor detection because this time frame is shorter than, and thereby falls entirely within, the sensor sampling and response period.
Therefore, we posit that the fall time is crucial in relation to the sensor's sampling frequency. 
Thus, we investigate the success of our approach in sensor bypass when altering the target device's clock frequency relative to a fixed voltage fall time. 
Note that a sensor's sampling rate may not be equal to the device clock rate, but rather is likely derived from the system clock rate. }

\begin{figure}[t]
    \centering
    \setlength{\abovecaptionskip}{0pt}
    \includegraphics[trim=0 250 10 250, clip, width=0.45\textwidth]{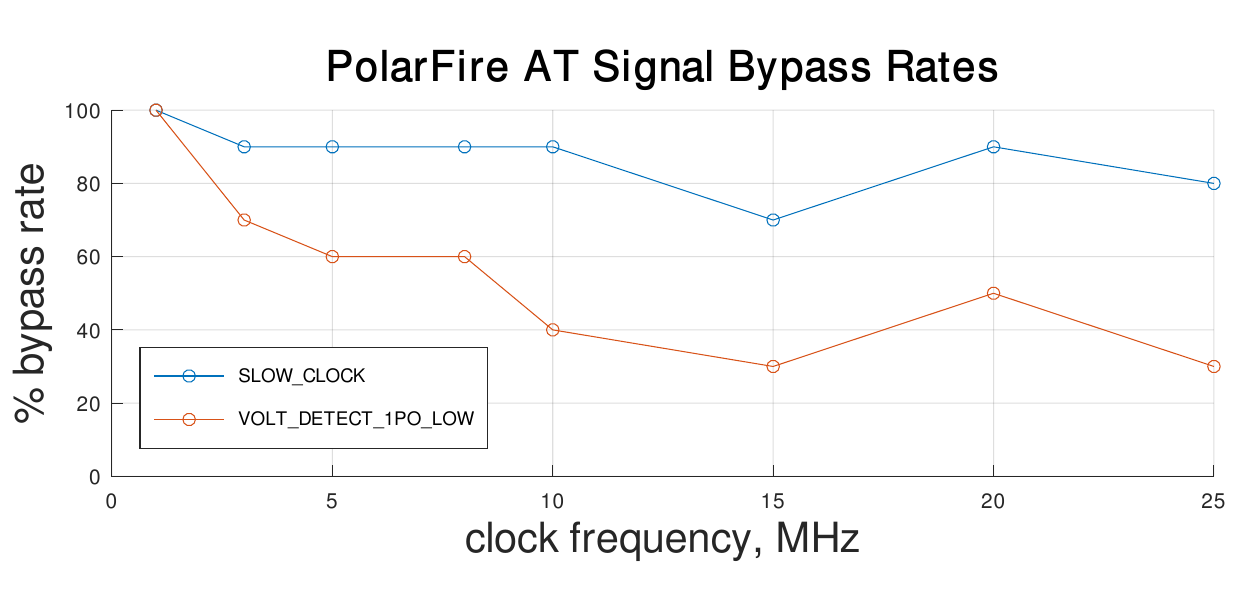}
    \caption{The success rate of bypassing the sensors using undervolting at various clock frequencies used by the anti-tamper module of  Microchip FPGA}
    \label{fig:voltage_clk}
\end{figure}

{\cref{fig:voltage_clk} shows the success rate of undervolting sensor bypass attempts against the Microchip anti-tamper module when setting the FPGA to various clock frequencies. 
We find that for our fixed fall time of 430\,ns, clock and voltage sensor bypass success rates are higher for lower clock frequencies, which matches our expectations.
Note that the fall time can be further decreased by deploying high-bandwidth instruments and high-frequency connections to the PCB.}

{
In the case of the AMD/Xilinx sensors, we find the clock frequency to have no impact on bypass success, which is 100\% for the same 
range of tested frequencies. 
This is even the case using the setup with the slower 80\,$\mu$s falling time. 
Since this is well below the sample rate used for the XADC, as discussed in \cref{sec:xilinx_xadc}, we expect the sensor's possible pipelined architecture to be responsible.}

\subsection{Defeating OpenTitan's Alert Handler}
\begin{figure}[t]
    \centering
    \includegraphics[width=0.49\textwidth]{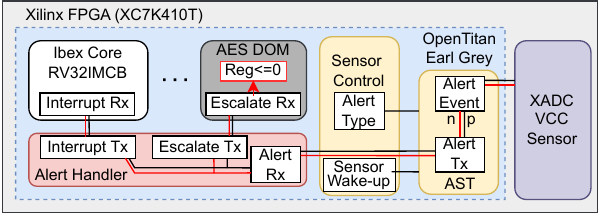}
    \caption{Architectures of OpenTitan Earl Gray Analog Sensors and Alert Handling Mechanism}
    \label{fig:ot_arch}
\end{figure}

OpenTitan is an open-source silicon root of trust (RoT) project that provides a secure, transparent, and verifiable foundation for computing systems. It implements side-channel protected cryptographic functions, secure boot, device identity, and runtime attestation, and is designed for high-assurance environments.
Earl Grey is the first full-chip implementation of the OpenTitan open-source RoT, designed as a low-power 32-bit RISC-V microprocessor. 
It features dedicated, secure cryptographic accelerators (e.g., HMAC and CSRNG), secure memory (flash, SRAM, OTP), and other hardened countermeasures, such as memory scrambling and enhanced physical memory protection~\cite{opentitanEG}. 
Earl Grey is equipped with various defenses, including detection-based countermeasures against side-channel and fault attacks.
\cref{fig:ot_arch} illustrates the high-level diagram of OpenTitan's Earl Grey analog sensing and response mechanism.
The \textit{ Analog Sensors Top} (AST)~\cite{opentitanAST} module of OpenTitan provides an interface to analog/digital elements that monitor key environmental parameters such as voltage, clock integrity, and temperature. 
These sensors detect anomalies, such as voltage or clock glitches, that may indicate tampering attempts. 
Moreover, the \textit{Sensor Controller} receives analog alert signals from the AST and forwards them to the alert handler, classifying each as recoverable or fatal. It also supports wake-up signaling, status readback~\cite{opentitanSC}.  
Although multiple analog sensors can be deployed and connected to OpenTitan, we focus on AMD/Xilinx's XADC voltage sensor (See \cref{sec:xilinx_xadc}). As shown, the Earl Grey facilitates differential signaling for its sensors to increase reliability and protect against fault attacks~\cite{opentitanAST}. Alert signals are then carried to the \textit{Alert Handler} and are classified, and initial interrupts to the processor are raised accordingly. If the processor fails to respond in time, the alert handler escalates these alerts through programmable hardware actions such as chip reset, key erasure, or privilege lowering to contain potential threats. Differential alert signaling and programmable escalation protocols across four severity levels are provided in this module~\cite{opentitan}. Although the escalation and response mechanism is handled by software by default, hardware/software transactions create \textit{hundreds of microseconds} of latency that can be bypassed easily by our rapid undervolting. In this case, we utilize a fast-track response in hardware via OpenTitan's alert handler, which can be enabled by setting the accumulation trigger to 1 and configuring escalation to begin at phase 0, allowing critical alerts to trigger a response in as few as \textit{4 clock cycles} when fully synchronous.

In this experiment, we follow the same procedure for deploying the XADC as described in \cref{sec:xilinx_xadc}. With a similar principle, a falling time of approximately 
80\,$\mu$s is fast enough to surpass the differential signaling of the sensor, regardless of the underlying mechanism provided by OpenTitan. Our experiments confirm that the key register (masked values) assignments indeed do not occur. Nevertheless, the minimum latency introduced by the OpenTitan in such a scenario can be estimated to be $4\times\frac{1s}{24Mhz} =166.66ns$, where the default peripheral domain clock in Earl Grey runs at $24Mhz$. In the case where the sensor latency is negligible, an undervolting attack with a falling edge faster than $166.66ns$ (which is feasible with the same experimental attacker setup) disables differential alert propagation in OpenTitan and renders the response circuitry ineffective.

\section{Side-channel Results}
\label{sec:results}


Having shown that undervolting can circumvent on-chip sensors to induce the conditions necessary for static SCA, namely an idle clock and continued data retention, we now investigate the viability of static SCA data extraction techniques against undervolted targets. 
Specifically, we use LLSI to perform direct register value readout, and we use IA to extract masked key shares from a cryptographic circuit.

\subsection{Hibernated LLSI Attack}\label{sec:llsi_attack}


For LLSI to be feasible in undervolted conditions we must be able to clearly distinguish between a register in the 0 and 1 state from scans. 
We prepare our target FPGA with a register, and to discern its physical location we configure it to toggle between 0 and 1 at a certain frequency.
This enables us to first localize it with LVI. 
Then, for optimal LLSI scans we used high laser power (90\%) with a sufficiently large laser wavelength $\lambda$\,=\,1300\,nm to avoid inducing bit flips, ensuring we inject no unintended faults. 

\begin{figure}[t]
    \setlength{\abovecaptionskip}{0pt}
    \setlength{\belowcaptionskip}{0pt} 

    \begin{subfigure}{0.22\textwidth}
        \centering
        \includegraphics[trim=0 0 0 100, clip, width=\textwidth]{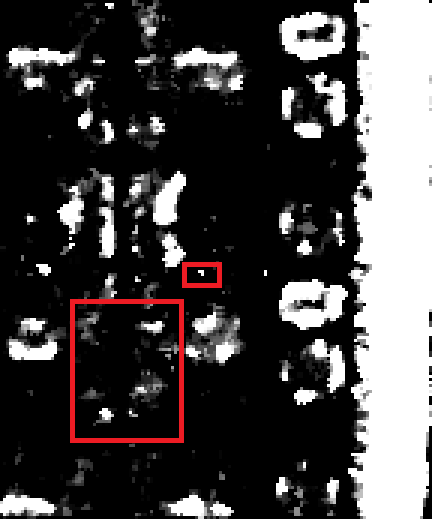}
        \caption{}
        \label{fig:llsi_h0p}
    \end{subfigure}
        \hfill
    \begin{subfigure}{0.22\textwidth}
        \centering
        \includegraphics[trim=0 0 0 100, clip, width=\textwidth]{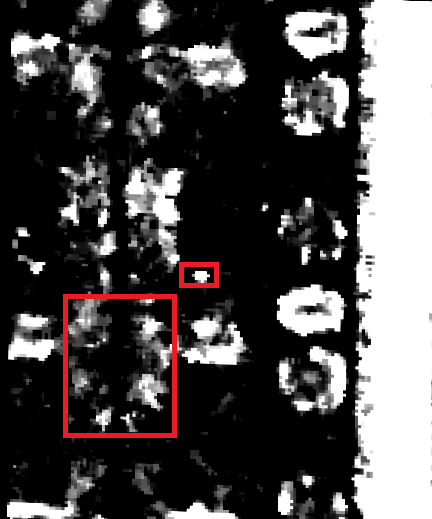}
        \caption{}
        \label{fig:llsi_h1p}
    \end{subfigure}
    \caption{LLSI images of a register in 0 (left) and 1 (right) states on a hibernated FPGA, major differences boxed}
    \label{fig:llsi_h}
\end{figure}
\begin{figure}[t]
    \setlength{\abovecaptionskip}{0pt}
    \setlength{\belowcaptionskip}{0pt} 
    
    \centering
    \begin{subfigure}{0.22\textwidth}
        \centering
        \includegraphics[trim=0 0 0 100, clip, width=\textwidth]{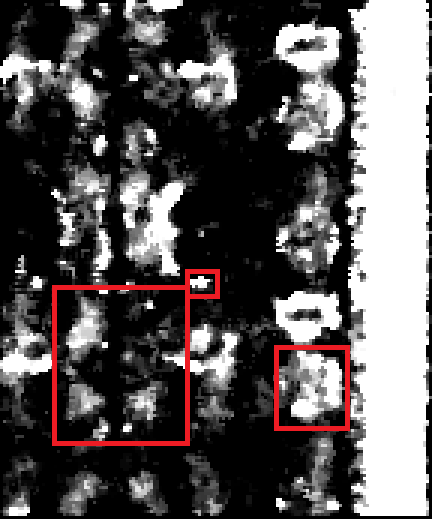}
        \caption{}
        \label{fig:llsi_d0}
    \end{subfigure}
    \hfill
    \begin{subfigure}{0.22\textwidth}
        \centering
        \includegraphics[trim=0 0 0 100, clip, width=\textwidth]{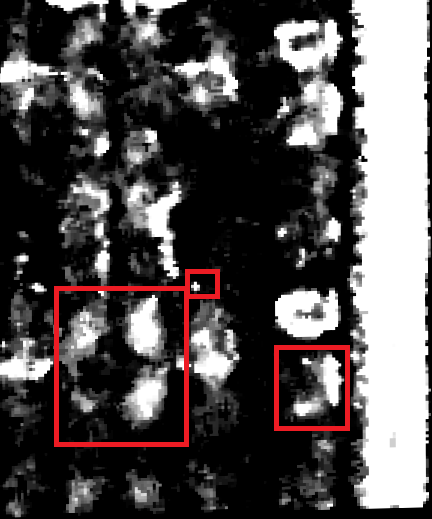}
        \caption{}
        \label{fig:llsi_d1}
    \end{subfigure}
    \caption{LLSI images of a register in 0 (left) and 1 (right) states on a non-hibernated FPGA, major differences boxed}
    \label{fig:llsi_d}
\end{figure}

We apply similar image processing techniques to our scans as those used in photon emission analysis~\cite{mehta_1/0_2024}.
Specifically, we 
apply a median filter to remove salt-and-pepper noise, followed by a bilateral filter to smooth the image while preserving edges.
The processed images, shown in \cref{fig:llsi_h0p,fig:llsi_h1p}, clearly distinguish between the two logic states.
For reference, we also compare this to LLSI images of the same register on the same target in normal (non-hibernated) conditions, see \cref{fig:llsi_d}. 
The contrast between 0 and 1 states is more distinct in nominal conditions. However, there is still sufficient distinguishability to directly extract data from hibernated targets using LLSI. 

\subsection{Hibernated Impedance Attack}\label{sec:ia_attack}

The practicality of IA against undervolted targets similarly depends on the distinguishability of individual bits. 
We perform a template impedance analysis attack~\cite{monfared2023leakyohm}, which first requires profiling to identify the specific frequency bands where the responses depend on data, localized to specific target registers. 

Our target is an AES protected with 3-share Domain-Oriented Masking (DOM)~\cite{gross2016domain}, used in OpenTitan. 
The attack scenario targets masked key bytes loaded into the internal key registers.
We also equip the target with the voltage sensor in \cref{sec:xilinx_xadc} and clock sensor protection from \cref{sec:BTdefeat} for these registers.


In this scenario, we attack when the AES key's first byte shares and the corresponding input byte shares are loaded into the target. 
This makes for 24 distinct bit-level profiling tasks (\( 8 \times 3 \) shares).
We perform undervolting at a hibernation voltage of \textit{0.64\,V} for both profiling and attack phase to disable the sensor. 
For profiling, we collect \( N_p = 20{,}000 \) traces, each contributing to profiling all target bits.
We separate $N_p$ traces into two classes of $k=0$ and $k=1$, where $k$ is the target bit to be profiled.
For a given bit we define the SNR at a frequency index \( f \) as:
\[
\text{SNR}(f) = \frac{\left( \mu_0(f) - \mu_1(f) \right)^2}{\frac{1}{2} \left( \sigma_0^2(f) + \sigma_1^2(f) \right) + \epsilon}
\]
where $\mu_0(f)$ and $\mu_1(f)$ are the mean values of class $(k=0)$ and class $(k=1)$ traces at frequency \( f \). \( \sigma^2 \) represents the variance, and \( \epsilon \) is a small constant added for numerical stability (e.g., \( \epsilon = 10^{-8} \)).
We extract the Points of Interest (POIs) in the frequency domain by picking those that maximize the SNR. The details of POI selection and parameterization is described in \cref{apx:ia_pr}.

\begin{figure}[t]
    \setlength{\abovecaptionskip}{0pt}
    \setlength{\belowcaptionskip}{0pt} 
    
    \centering
    \begin{subfigure}{0.23\textwidth}
        \setlength{\abovecaptionskip}{0pt}
        \centering
        \includegraphics[width=\textwidth]{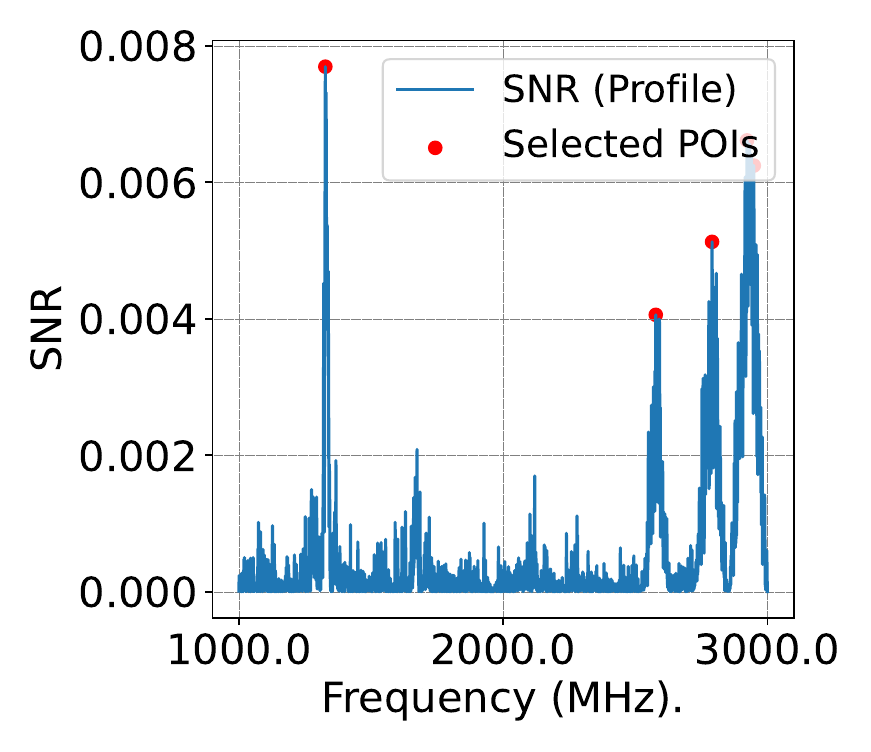}
        \caption{}
        \label{fig:IA_SNR_bit1}
    \end{subfigure}
    \hfill
    \begin{subfigure}{0.23\textwidth}
        \setlength{\abovecaptionskip}{0pt}
        \centering
        \includegraphics[width=\textwidth]{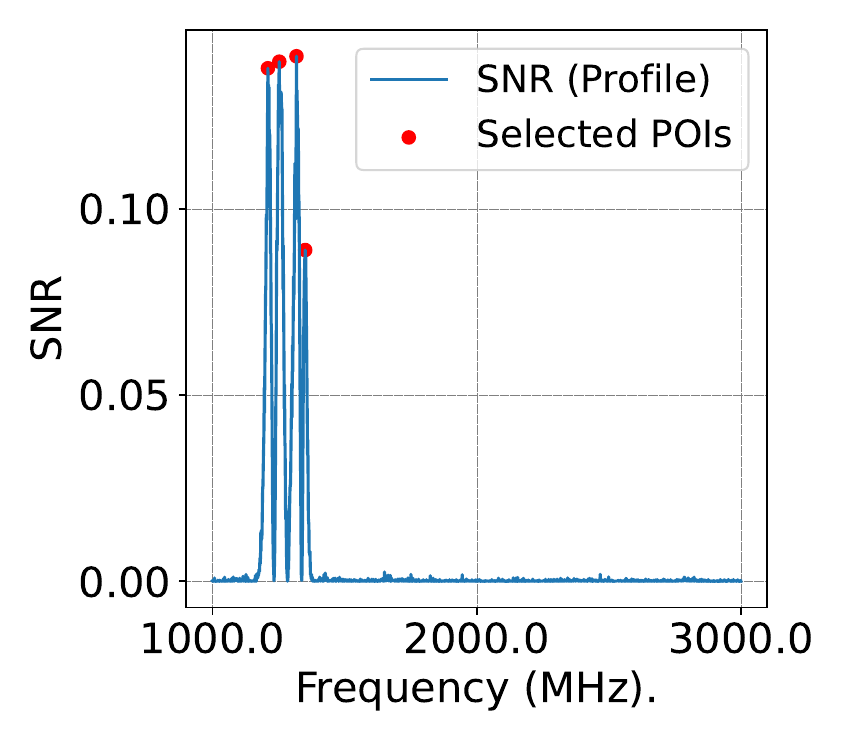}
        \caption{}
        \label{fig:IA_SNR_bit4}
    \end{subfigure}
    \caption{SNR curve for \texttt{Bit=1}(a) and \texttt{Bit=4}(b) of the key share across different frequencies. POIs shown in red.}
    \label{fig:IA_SNR_bit}
\end{figure}

\begin{figure}[t]
    \centering
    \includegraphics[width=0.45\textwidth]{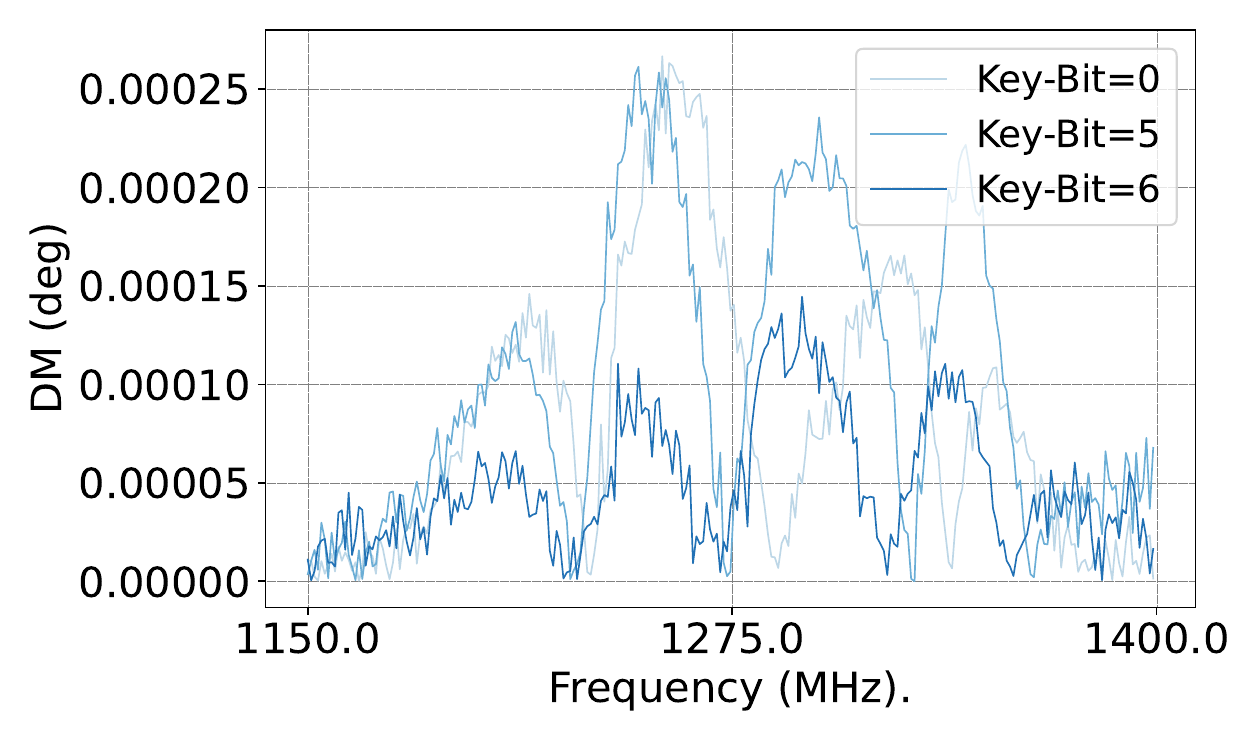}
    \caption{Phase Difference of Means impedance leakage for \texttt{Bit=0}, \texttt{Bit=5}, and \texttt{Bit=6} in one share byte over a zoomed in frequency window.} 
    \label{fig:IA_inter_bit}
\end{figure}

\cref{fig:IA_SNR_bit1,fig:IA_SNR_bit4} illustrate the observed bit-level leakage across the frequency spectrum for the \texttt{KeyBit=1}, \texttt{KeyBit=4} from the first share byte.
These results are visualized using the SNR, emphasizing that distinct POIs across frequencies for individual bits enable template attacks to isolate and extract bit-specific information effectively. On the other hand, a similar analysis can be done using the phase Difference of Mean (DM) metric. For instance \cref{fig:IA_inter_bit}, illustrates the $DM=\mu_0(\phi_f) - \mu_1(\phi_f)$ for $k\in\{0,5,6\}$. zoomed in a specific frequency window, illustrating the distinct bit-level impedance leakage. 

Following the profiling phase, we conduct a single-trace attack against the instance with unknown key shares. To mitigate noise and improve robustness, we perform VNA-enabled averaging using \( N_{\text{avg}} = 400 \) repeated acquisitions for the same attack trace.
The impedance template attack successfully recovers all individual bits from each share when enough iteration of averaging is performed.
Naturally, the recovered bits from the three shares can then be combined to reconstruct the full first byte of the AES master key. For a detail key bit extraction analysis see \cref{apx:ia}.

\section{Undervolting-Resilient Countermeasure}
\label{sec:newBT}

This section proposes an improved design of the clock sensor systems evaluated in \cref{sec:defeat} that equips them to handle our undervolting attacks. 
We implement the updated design in both the AMD/Xilinx and Microchip PolarFire SoC FPGAs and find them to protect against our attacks. 

To recap, both original clock sensor variants assert an alarm signal upon detecting a stopped clock, but undervolting hinders their subsequent response actions for performing a masked clear. 
Specifically, the mechanism for synchronous latching, to overwrite sensitive values with randomness, is affected and cannot complete.
To remedy the problem, we propose modifying the masked clear mechanism. 
We observe that when using flip-flops that support asynchronous resets, such resets complete correctly, even in a brownout state.
This robustness stems from these reset paths typically being built to remain dependable even when the device operates at, or marginally beyond, its process, voltage, and temperature limits.
However, a naive use of asynchronous resets tied directly to the alarm signal would zeroize registers.
As~\citet{dumitru2025borrowed} point out, this would directly leak their Hamming weight dynamically.
A potential fix is also to use flip-flop primitives, which instead asynchronously \emph{preset} to 1.
We verify that such flip-flops are supported and that the preset mechanism also works in brownout conditions.
However, whether a flip-flop resets or presets is typically fixed to the primitive type, i.e., the same flip-flop cannot be dynamically configured to work as either type. 
Hence, directly using presets instead would still result in leakage.

To avoid introducing dynamic leakage upon an asynchronous clear, we can randomize the overall number of transitions by constructing unbiased registers based on a combination of these flip-flop primitives. 
Our scheme involves randomly selecting which type of register primitive is used to store each bit of a sensitive value. 
Since the primitives are fixed components that require instantiation at build time, for the storage of a single sensitive bit, we implement one flip-flop of each kind and use randomness to dynamically select between them at runtime. 
Moreover, we also want to ensure that among the random number of transitions upon a clear, there is an equal number of 0\,$\rightarrow$\,1 as 1\,$\rightarrow$\,0 transitions so as to avoid information leakage due to imbalances. 
To that aim we propose a \emph{complementary register} design that enables performing masked clears using the asynchronous reset/preset mechanisms. 

\begin{figure}[t]
    \centering
    \includegraphics[width=0.4\textwidth]{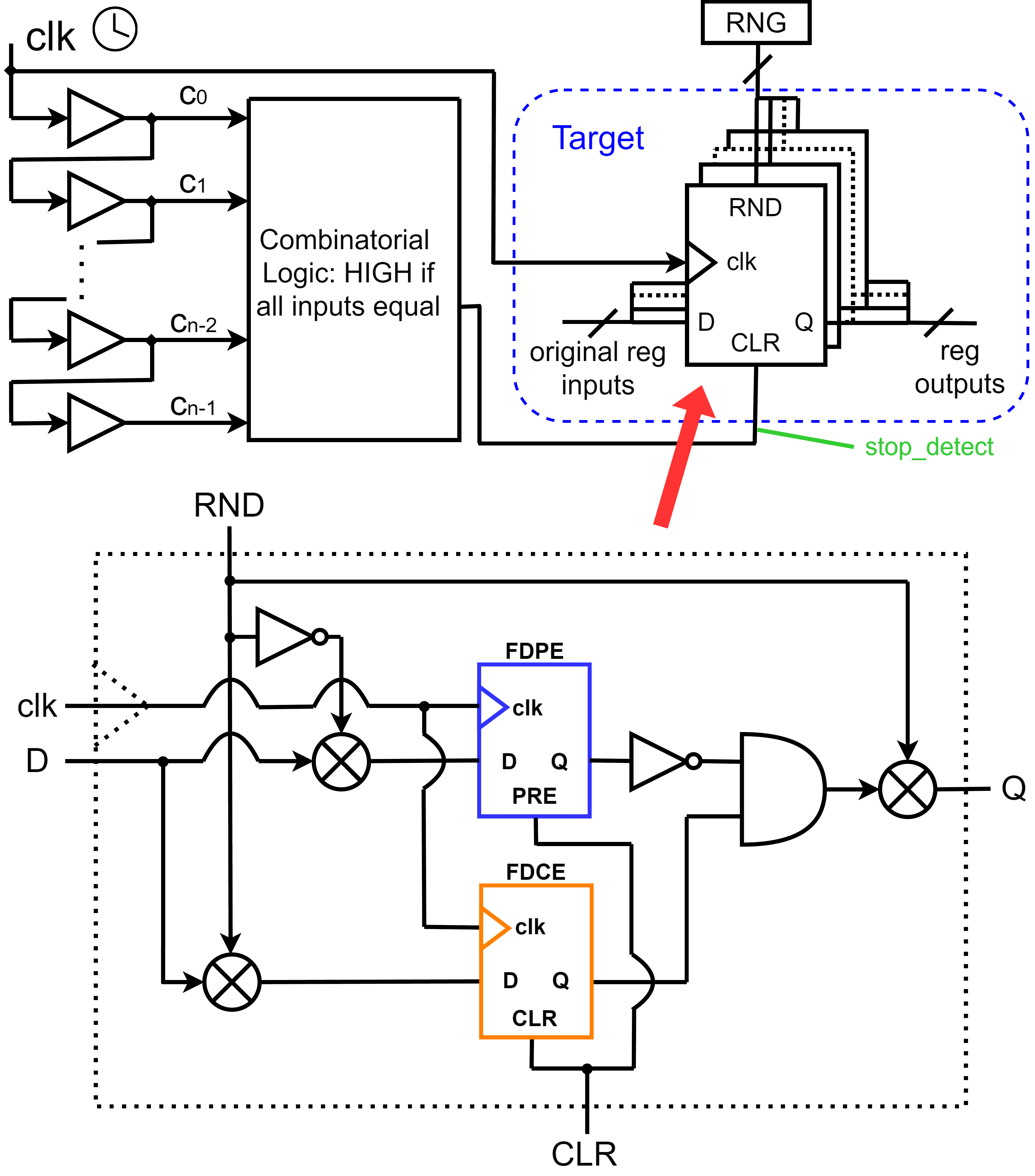}
    \caption{Complementary Registers within asynchronous delay-based Borrowed Time countermeasure with asynchronous clearing, equipped to handle undervolting attacks.} 
    \label{fig:CMR}
\end{figure}

\parhead{Design and Implementations.\footnote{The source code for our design artifacts is available at \hbox{\url{https://github.com/0xADE1A1DE/Borrowed-Time}}}}
We implement the revised clock sensor systems into both the Xilinx/AMD and Microchip PolarFire SoC FPGA platforms attacked (with sensor bypasses) in \cref{sec:defeat} and \cref{sec:results}. 

\cref{fig:CMR} depicts the revised asynchronous delay-based clock sensor system that we implement in the 7\nobreakdash-Series AMD/Xilinx FPGAs~\cite{7seriesug} incorporating complementary registers.
To the outside, they work as a regular single-cycle, rising-edge-triggered registers. 
Internally, they comprise two flip-flops with different properties: one which can be asynchronously reset to~0, and another which can be asynchronously preset to~1. 
Respectively, we implement these using FDCE (orange) and FDPE (blue) primitives. 
Both registers store the sensitive data in some form. 
For each bit of sensitive data, one flip-flop  stores the bit and the other stores its complement. 
Which of the two flip-flops stores the data itself, i.e., the polarity of the complementary register, is determined by a random selector bit. 
The same ``clear" signal sourced from the clock sensor alarm is connected to both the asynchronous reset and preset lines.

The sources of randomness used with our countermeasure 
are also likely to be affected by undervolting.
However, this does not impact our countermeasure since we ``pre-load'' the complementary registers with randomness generated in previous clock cycles during normal circuit operation. 

We extend the overhead analysis from the original work~\cite{dumitru2025borrowed}, finding the cost of our revised design to be similar and modest overall. 
As in~\cite{dumitru2025borrowed}, \cref{table:overhead} reports power and utilization overheads with respect to a lightweight AES circuit~\cite{BorrowedTime_github} (Base), the circuit equipped with the original Borrowed Time clock sensor design (BT\textsubscript{orig}), and our updated version with complementary registers (BT\textsubscript{CR}). 
The randomness requirements are unchanged. 
With the updated design we double up the number of flip-flops for sensitive values and use additional LUTs for each complementary register's input and output logic (shown as logic gates in \cref{fig:CMR}). 
Conversely, LUTs used in the edge generation and data multiplexing circuitry from BT\textsubscript{orig} are no longer required. 
While complementary registers impose no overhead in terms of extra clock cycles, if they are built into a design's critical path, the extra LUTs may affect the circuit’s maximum possible clock frequency. 
This is the case in our evaluated AES circuit, where the critical path is slightly longer than the original Borrowed Time countermeasure.  BT\textsubscript{orig} instead only added a data multiplexing stage at the register inputs. 
The updated design also cuts down in other areas not captured by the table as we no longer require the same clock multiplexing and buffering circuitry. 
We further note that our protection (along with its overhead), when incorporated within a target with masking our protection, need only apply to one share of sensitive values. 

\begin{table}[ht]
\setlength{\belowcaptionskip}{0pt} 
  \small
\centering
  \begin{tabular}{@{}lccccc@{}}
 &  \textbf{Power [mW]} & \textbf{c.p. [ns]} &
 \textbf{f\textsubscript{max} [MHz]} & \textbf{LUTs} & \textbf{Regs} \\

\toprule 
Base & 116 & 5.080 &
196.9 & 1387 & 535 \\
 BT\textsubscript{orig} & 138 & 6.009 & 
 166.4 & 4079 & 1163 \\ 
 BT\textsubscript{CR} & 143 & 6.628 & 
 150.9 & 4075 & 1294\\
\bottomrule
\end{tabular}
  \caption{Clock sensor countermeasure overheads on lightweight base AES core: power, critical path (c.p.) with corresponding max frequency, LUT and Register utilization.}
\label{table:overhead} 
\end{table}

In the Microchip PolarFire SoC FPGA we implement the PLL-based clock sensor with similar architecture to \cref{fig:CMR}, simply swapping out the detector circuit for a PLL contained in a PolarFire Clock Conditioning Circuitry (PF\_CCC) module and its output \texttt{PLL\_LOCK} in place of \texttt{stop\_detect}. 
In place of FDPE and FDCE we use equivalent PolarFire flip-flop macros~\cite{polarfiremacros}, DFN1E1C0 and DFN1E1P0, which asynchronously clear and preset, respectively. 
Both primitives have active low reset/preset, so we tie \texttt{PLL\_LOCK} to them directly without negating it. 

We validate the full improved countermeasure designs on both target FPGA families in the context of our undervolting attacks and find them to protect against the attacks.

\section{Discussion}\label{sec:discussion}

\subsection{Applicability to Static Power SCA}
The main conditions necessary for static power SCA are sensitive data retention in the target during an idle clocking period, which we show to be simultaneously achievable with Chypnosis.
We expect the data extraction methods of these attacks to still be possible with undervolting, however it is likely to affect their performance. 
Previous works~\cite{moos2019static,cassiers2023prime} point out that increased supply voltage amplifies leakage, but also that control over this parameter is not strictly necessary and only serves to reduce the overall number of traces needed. 
Moreover, Chypnosis does not preclude controlling any of the other parameters that maximize attack performance such as high temperature, intra-trace averaging, and post-processing. 
Importantly, Chypnosis relaxes the requisite adversary capabilities to not need any form of controllable clock, and to work in the presence of clock and voltage sensors. 
We leave investigation of undervolting static power SCA attacks to future work. 
We reaffirm that the revised clock sensor countermeasure we propose in \cref{sec:newBT} offers resilience to these attacks.


\subsection{Comparison with Voltage Glitching Attacks}
The undervolting used in our proposed attack might initially appear similar to conventional voltage glitching attacks~\cite{tang2017clkscrew,buhren2021one}. In voltage glitching, the adversary induces a transient voltage drop to cause timing violations in sequential logic, potentially triggering unauthorized state transitions. Such faults could, in some cases, bypass the response mechanisms of the countermeasures discussed in this paper.
However, unlike glitching attacks, where the voltage returns to its nominal level and the system resumes normal operation, our approach involves a permanent voltage drop and clock halt, which is crucial for static SCA.
Moreover, in glitching scenarios, various countermeasures such as Error Correction Codes (ECC)~\cite{spensky2021glitching} can be employed to protect against transient faults. 
In contrast, the persistent nature of the voltage drop in our attack disables on-chip fault detection and response mechanisms, rendering them ineffective.


\subsection{Comparison with Data Remanence Attacks}
At first glance, one might assume that our proposed attack is similar to data remanence~\cite{oren2013effectiveness,anagnostopoulos2018low}, Cold Boot~\cite{halderman2009lest} or Pentimento~\cite{drewes2024pentimento} attacks, in which the adversary exploits charge retention or bias temperature instability effects in underlying transistors to recover data previously stored in memory. However, our attack differs in several key ways.
First, it does not rely on temperature effects for data recovery. Additionally, those attacks generally assume that the adversary can run their firmware or bitstream on the chip after the sensitive data has been wiped, exploiting analog features of the memory (e.g., SRAM metastability or flip-flop propagation delay) to recover the contents. In contrast, our approach directly measures memory content that remains retained during a brownout condition, with the assumption that the adversary cannot take control of the chip by executing code or reading back any data at a later time.


\subsection{Applicability to ASICs}
Two questions may arise regarding the applicability of our attack to ASICs:
(1) Can this technique inhibit the overwriting of secrets by voltage and clock sensors implemented in ASICs? (2) Can it also stop ASIC clocks?
For the first question, it is important to note that the voltage and clock sensors in the Microchip SoC, as well as the voltage sensor in AMD/Xilinx FPGAs, are implemented as ASICs (hard IP). Our results clearly demonstrate that these sensors can be bypassed.
Regarding the second question, prior work has already demonstrated that brownout conditions exist for SRAMs on ASICs and microcontrollers~\cite{nedospasov2013invasive}. 
However, FPGAs have inherently larger capacitances than ASICs, which may make them more vulnerable to this attack compared to ASICs.
As a result, while a detailed comparison between FPGAs and ASICs is beyond the scope of this work, our findings strongly indicate that the core mechanisms exploited in our attack are not exclusive to FPGAs and warrant further investigation in future work.

\subsection{End-to-end Key Extraction}

{{In \cref{sec:llsi_attack} and \cref{sec:ia_attack} we demonstrated register bit extraction and a masked AES key-byte extraction on an undervolted FPGA using impedance and LLSI attacks, respectively.
To perform complete key recovery, an attacker can simply repeat the same procedure for all key bytes over a longer period.
Since undervolting halts the clock while retaining data indefinitely, a longer measurement time for extraction has no inhibitory consequences for the approach.}}
\section{Conclusion}\label{sec:conclusion}
In this work, we introduced \emph{Chypnosis}, a novel undervolting attack that exploits the vulnerability of chips during brownout conditions. 
By inducing rapid voltage drops, we demonstrated that it is possible to halt all internal clock sources and freeze the circuit's state without triggering conventional clock, voltage, or brownout detection (BOD) sensors, and consequently, the erasure of sensitive data. 
This enables adversaries to extract the retained secrets in flip-flops and other non-volatile memories using static side-channels, such as LLSI and IA.
Our extensive experiments on SRAM-based and Flash-based FPGAs validated our claims.
We also showed that our attack can bypass the OpenTitan RoT's alert handler, demonstrating its real-world impact.
To mitigate the threats posed by Chypnosis, we proposed a revised clock sensor countermeasure design, which we demonstrated can protect all evaluated systems, even in brownout conditions.  


\section{Responsible Disclosure}\label{sec:disclosure}
Following the discovery of the vulnerability, we responsibly disclosed it to AMD, Microchip, and OpenTitan on June 7, 2025, upon completing the initial version of the manuscript. 
All parties have responded to initial contact and have remained in contact since the paper report was sent out immediately following their initial responses.
All parties acknowledged receiving the report.
Since the report's acknowledgments, we have held meetings with representatives from all relevant parties to discuss plans for addressing the issues and the embargo timelines. In each of the meetings, we informed the parties that the findings would be embargoed until at least September 4. AMD and Microchip publicly acknowledged the vulnerability~\cite{AMD-SB-8018,PSIRT-118}.

\section*{Acknowledgments}
This effort was sponsored by NSF Grants CNS-2150123 and CNS-2338069; Draper Scholars Program; Research and Development (R\&D) grant from the Massachusetts Technology Collaborative; an ARC Discovery Project number DP210102670; and the Deutsche Forschungsgemeinschaft (DFG, German Research Foundation) under Germany's Excellence Strategy - EXC 2092 CASA - 390781972. 

\bibliographystyle{IEEEtranSN}
\bibliography{ref_fixed}

\appendices
\crefalias{section}{appendix}
\section{Hibernation Characterization Algorithm}\label{apx:alg1}

\begin{algorithm}

\begin{algorithmic}

\Function{Hibernation Scan}{}

\For{$f \in \text{linspace}(f_{low}, f_{high}, f_{step})$} \Comment{Sweep frequency}
\For{$v \in \text{linspace}(V_{high}, V_{low}, -V_{step})$} \Comment{Sweep voltage}

    \State Write $(\text{PLL\_Freq}) \gets f$
    \State \Call{Debug\_Reg\_Reset()}{}
\Comment{Reset all values}

    \State \textit{/* Set Initial Delay */}
    \State Write $(\text{REG\_INIT\_DELAY}) \gets t_d$
    
    \State \textit{/* Set Test Duration */}
    \State Write $(\text{REG\_EVAL\_TIME}) \gets t_t$ \Comment{Minimum 0.5s}

    \State \textit{/* Trigger Evaluation */}
    \State Write $(\text{REG\_EXEC\_TEST}) \gets 0x01$
    \begin{tcolorbox}[colback=red!5!white,colframe=red!75!black]

    \State \textit{/* UNDERVOLTING */}
        \State Write $(\text{VAL\_Voltage}) \gets v$
    \State Wait $t=0.8s$ \Comment{Eval time $t>t_d+t_t$}
    
    \State \textit{/* Recovery phase */}
        \State Write $(\text{VAL\_Voltage}) \gets V_{high}$
            \State \textit{/* Reading debug regs */}
            
    \State $R_{rec} \gets$ Read(REG\_DEBUG\_VAL)
    \State $reg\_assign \gets R_{rec}[0]$
    \State $clock\_count \gets R_{rec}[1{:}10]$
    \end{tcolorbox}

    \State \textit{/* Crash Recovery */}
    \If{\Call{Debug\_Reg\_Reset()}{} $\neq 0$}
        \State $crash \gets \text{True}$
        \State \Call{Reprogram\_FPGA()}{}
    \Else
        \State $crash \gets \text{False}$
    \EndIf

    \State \textbf{Store} $\{f, v, reg\_assign, clock\_count, crash\}$

\EndFor
\EndFor

\EndFunction

\Function{Debug\_Reg\_Reset()}{}
    \State Write $(\text{REG\_DEBUG\_RST}) \gets 0x01$
    \State Wait 0.1 sec \Comment{Wiping registers}
    \State Write $(\text{REG\_DEBUG\_RST}) \gets 0x00$
    \State $R \gets$ Read(REG\_DEBUG\_VAL)
    \If{$R[0] \neq R_{Predefined}$}
        \State \Return $-1$ \Comment{Crash Detected}
    \Else
        \State \Return $0$
    \EndIf
\EndFunction

\end{algorithmic}
\caption{Undervolting Voltage-Frequency Scan}
\label{alg:hibernation-scan}

\end{algorithm}

\cref{alg:hibernation-scan} outlines in detail the scanning procedure presented in \cref{sec:characterization} for discovering the combined voltage and clock frequency hibernation threshold of a device. 
Our tests capture the following modes of failure:

\begin{itemize}[nosep,left=0pt]
    \item \textbf{Register Assignment Failure:} If the register assignment output differs from its expected value (e.g., \texttt{reg\_assign}\,$\neq$\,\texttt{0x88}), it indicates that flip-flops failed to latch the input due to timing violations introduced by low voltage.

    \item \textbf{Clocking Failure:} If the \texttt{clock\_count} remains near zero, it implies that the clock network failed to propagate or that the counter logic ceased functioning.

    \item \textbf{Crash State:} If the debug registers fail to return to a known state after reset, this is treated as a system-wide logic failure. It often correlates with unstable supply levels that corrupt control paths or flip-flop states.
\end{itemize}

\section{Impedance Attack Key Extraction Analysis}\label{apx:ia}

\begin{figure}[t]
    \setlength{\abovecaptionskip}{0pt}
    \setlength{\belowcaptionskip}{0pt} 
    
    \centering
    \begin{subfigure}{0.23\textwidth}
        \setlength{\abovecaptionskip}{0pt}
        \centering
        \includegraphics[width=\textwidth]{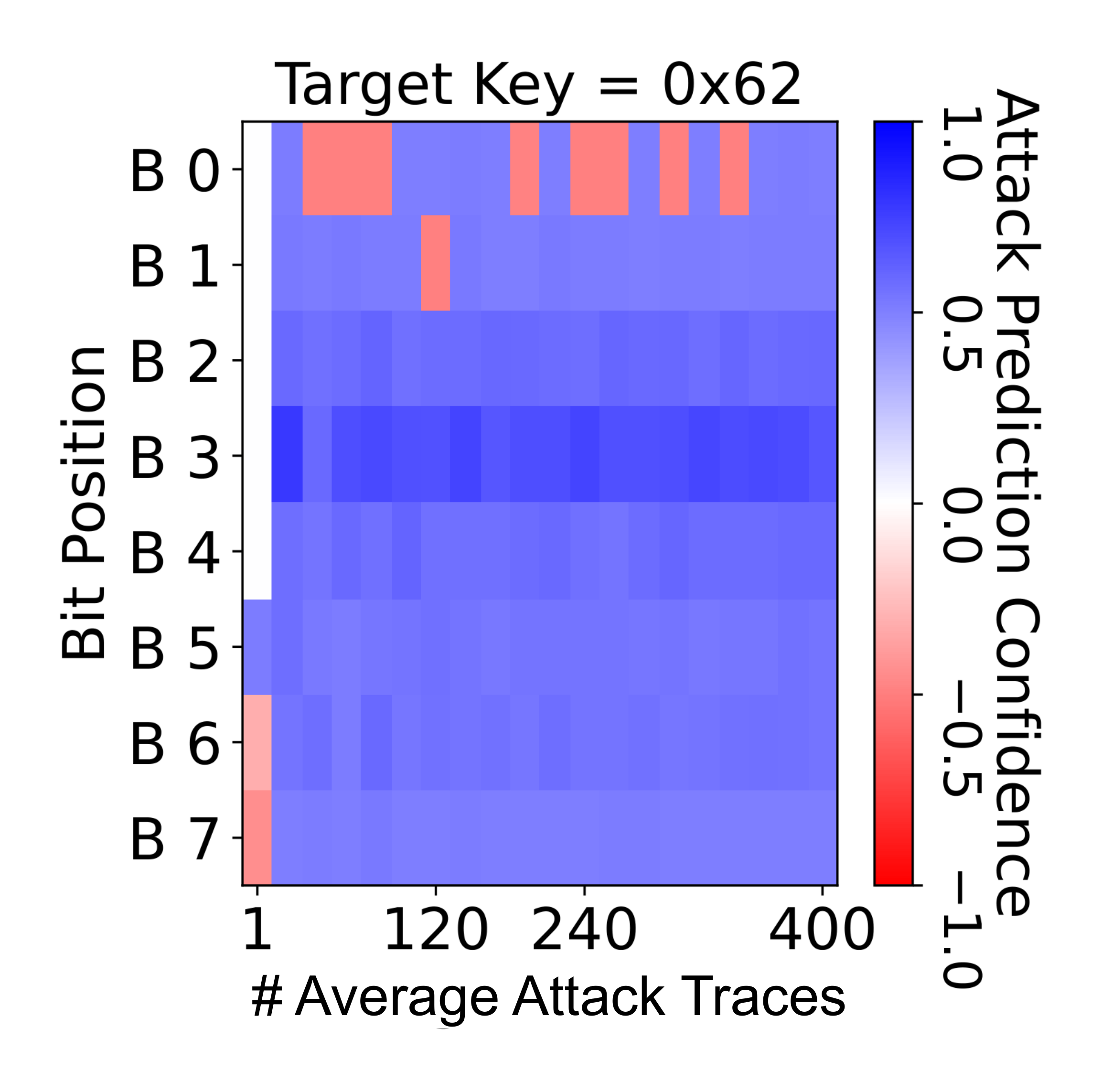}
        \caption{}
        \label{fig:LDA_62}
    \end{subfigure}
    \hfill
    \begin{subfigure}{0.23\textwidth}
        \setlength{\abovecaptionskip}{0pt}
        \centering
        \includegraphics[width=\textwidth]{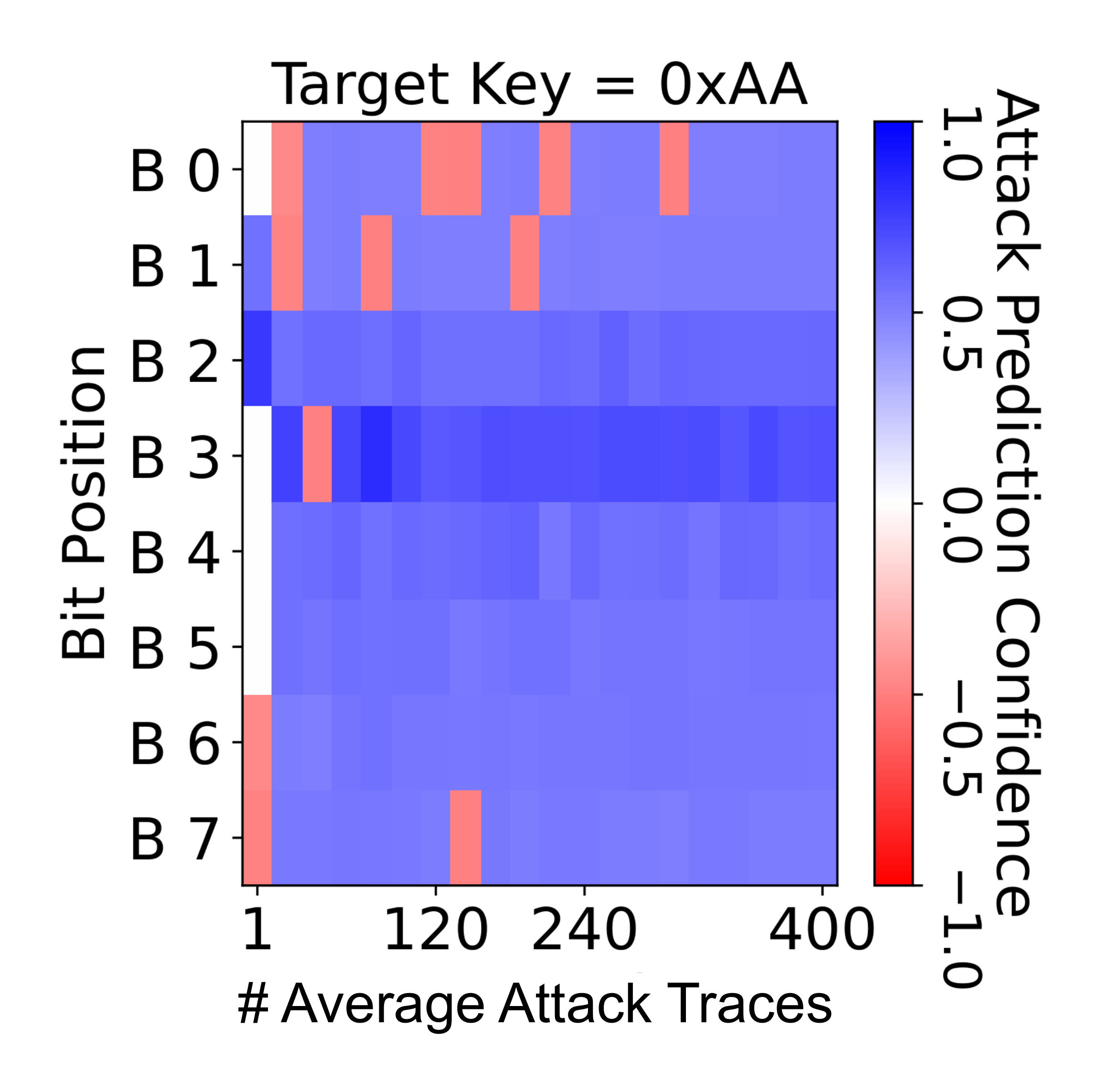}
        \caption{}
        \label{fig:LDA_aa}
    \end{subfigure}
    \caption{Attack prediction confidence via LDA method for a target with \texttt{KeyByte=0x62}(a) and \texttt{KeyByte=0xAA}(b). Blue and Red spots indicate correct and wrong predictions.}
    \label{fig:LDA}
\end{figure}

\begin{figure}[t]
    \setlength{\abovecaptionskip}{0pt}
    \setlength{\belowcaptionskip}{0pt} 
    
    \centering
    \begin{subfigure}{0.23\textwidth}
        \setlength{\abovecaptionskip}{0pt}
        \centering
        \includegraphics[width=\textwidth]{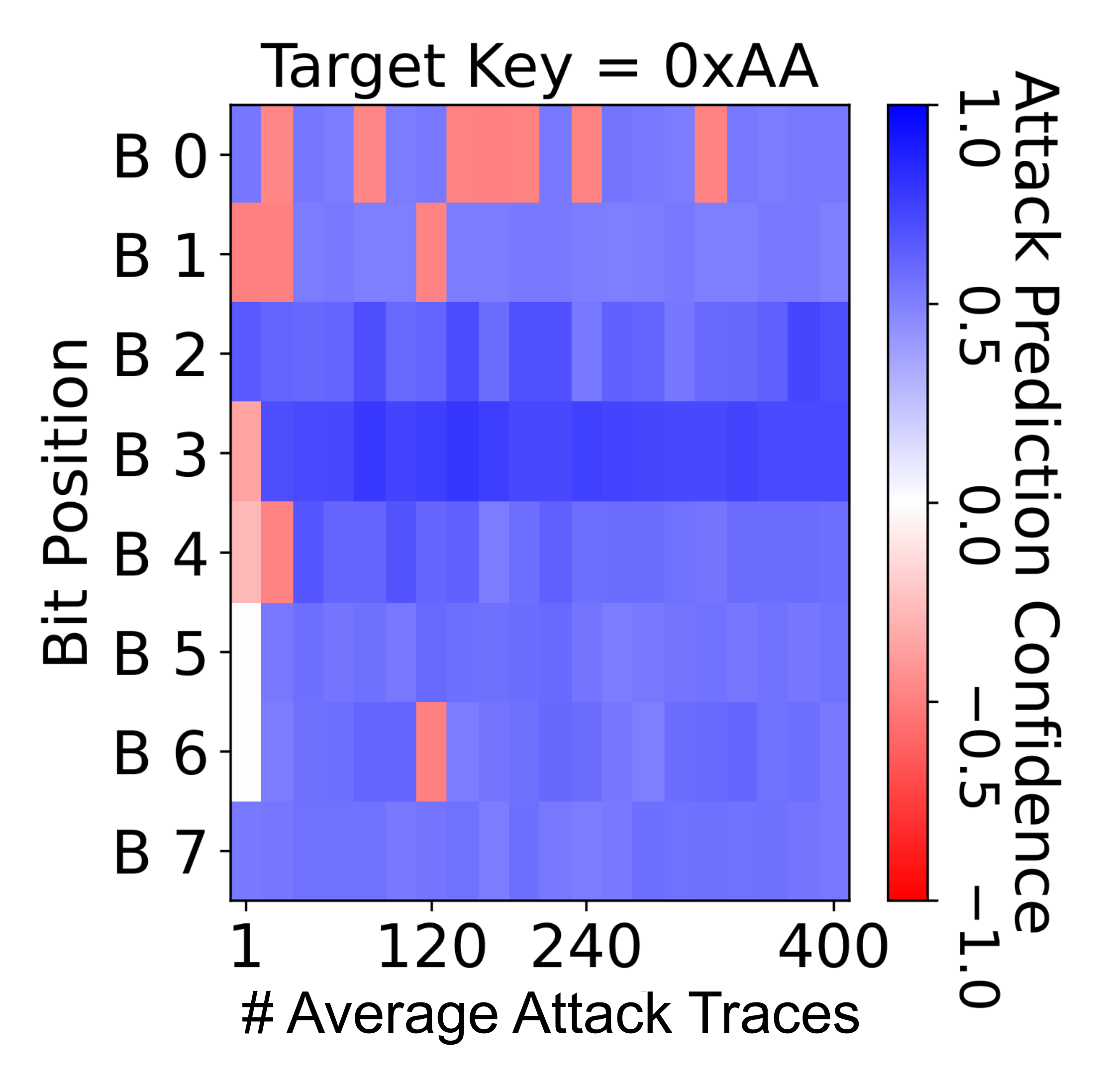}
        \caption{}
        \label{fig:RF_aa}
    \end{subfigure}
    \hfill
    \begin{subfigure}{0.23\textwidth}
        \setlength{\abovecaptionskip}{0pt}
        \centering
        \includegraphics[width=\textwidth]{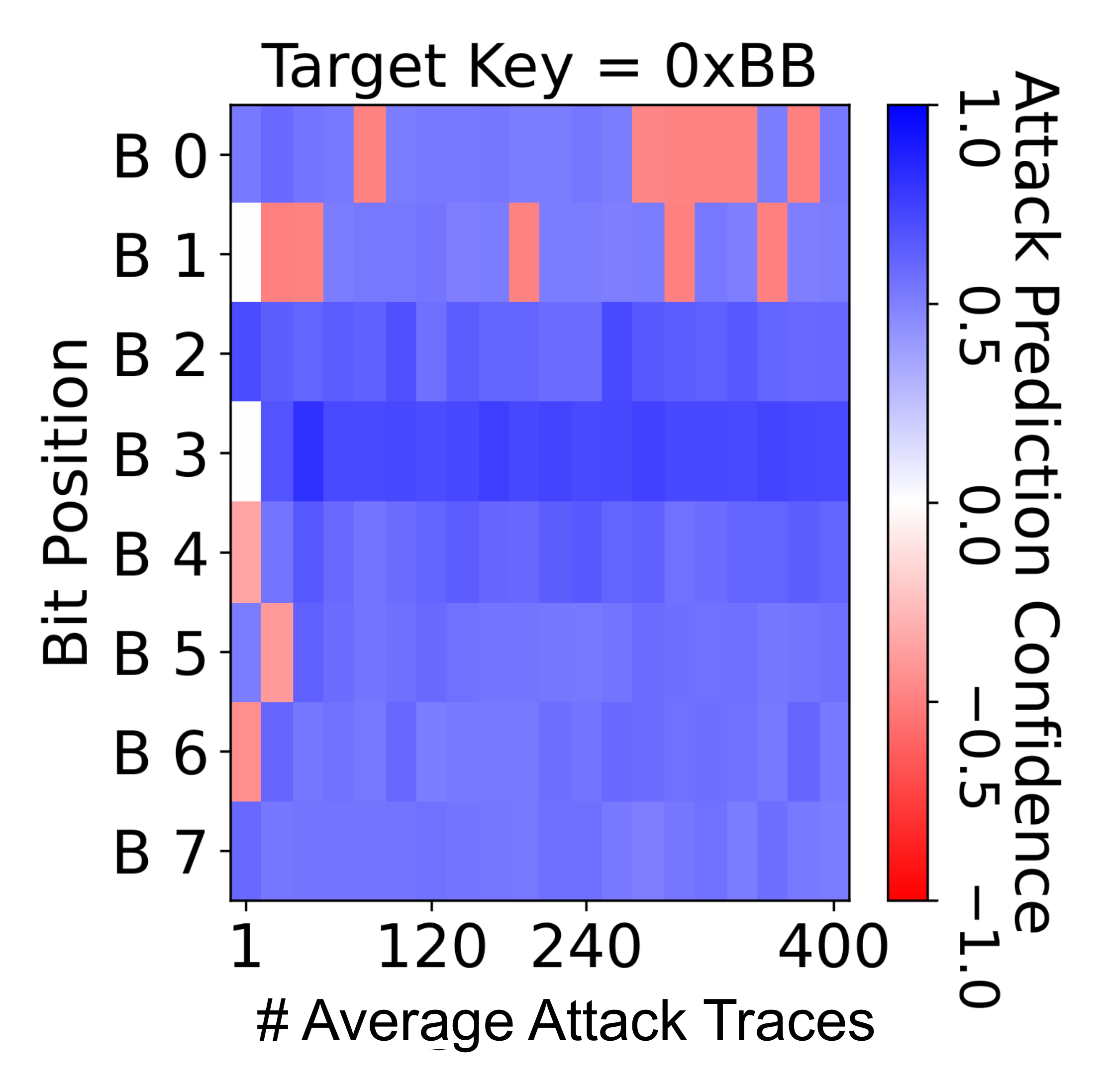}
        \caption{}
        \label{fig:RF_bb}
    \end{subfigure}
    \caption{Attack prediction confidence via RF method for a target with \texttt{KeyByte=0xAA}(a) and \texttt{KeyByte=0xBB}(b). Blue and Red spots indicate correct and wrong predictions.}
    \label{fig:RF}
\end{figure}

\cref{sec:ia_attack} describes the results of our Impedance Analysis attacks with Chypnosis.
Here we show examples of the post-attack key bit extraction, based on template matching scores \cref{fig:LDA,fig:RF} via LDA and RF methods, respectively.
These figures present the confidence of the attack prediction.
Blue colors represent the correct value, whereas red spots highlight the wrong prediction. 
In \cref{fig:LDA_62}, a trace with  \texttt{KeyByte=0x62} is captured and analyzed.
As the number of averaging increases, the template model tends to make fewer errors predicting the right value for key bits.
Furthermore, as shown in \cref{fig:RF_bb}, the RF model performs poorly predicting some bits (e.g., \texttt{Bit=0}) with a small amount of averaging.

\section{POI Selection in Impedance Attack Profiling }\label{apx:ia_pr}





During our profiling in \cref{sec:ia_attack}, we select the frequencies (POIs) that carry the most side-channel information. To choose the top-k POIs, we use empirically tuned parameters to constrain the candidate set $S(f)\subset SNR(f)$ with a \textit{Minimum Height} where $S(f) \geq \alpha \cdot \max(S)$ with a relative height threshold \( \alpha = 0.3 \), and enforce a \textit{Minimum Distance}. Hence, POIs are at least \( d_{\min} \) samples apart:
 
 \[
 |f_i - f_j| \geq d_{\min} \quad \forall i \neq j
 \]

\section{Clock Sensor System Timing}\label{sec:timingdiagram}
\begin{figure}[t]
    \centering
    \includegraphics[width=0.32\textwidth]{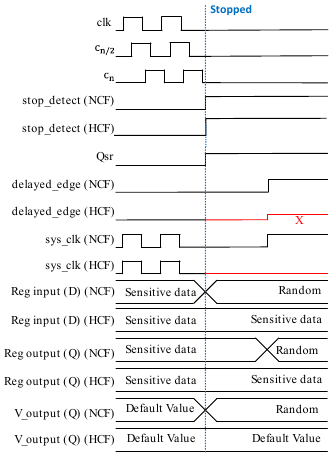}
    \caption{Timing diagram of the asynchronous circuit for clock halt detection and attempted register overwrite. Undervolting prevents the register from overwriting its content, causing it to retain the data. X indicates an unknown signal value during this process. NCF and HCF refer to normal clock freezing and hibernated clock freezing, respectively.} 
    \label{fig:BT_Timing}
\end{figure}

The timing diagram of the clock sensor (\cref{sec:BTdefeat}) is shown under attack~\cref{fig:BT_Timing}.


\end{document}